\documentclass[iop,apj]{emulateapj}
\usepackage{epsfig}
\usepackage{amsmath}
\usepackage{rotating}
\newcommand{\mnref}[1]{\hangindent=0.5in \hangafter=1 #1 \par}
\newenvironment{refs}{\parindent=0pt}{\parindent=1.5em}

\newcommand{\Msolar}{\mbox{\,$\rm M_{\odot}$}}
\newcommand{\Lsolar}{\mbox{\,$\rm L_{\odot}$}}
\def\gs{\mathrel{\raise1.16pt\hbox{$>$}\kern-7.0pt
\lower3.06pt\hbox{{$\scriptstyle \sim$}}}}
\def\ls{\mathrel{\raise1.16pt\hbox{$<$}\kern-7.0pt
\lower3.06pt\hbox{{$\scriptstyle \sim$}}}}

\begin{document}

\title
{The Red MSX Source Survey: the Massive Young Stellar Population of our Galaxy}
\author{S.L. Lumsden\altaffilmark{1}, 
M.G. Hoare\altaffilmark{1}, J.S. Urquhart\altaffilmark{2}, 
R.D. Oudmaijer\altaffilmark{1}, 
 B. Davies\altaffilmark{3},
  J.C. Mottram\altaffilmark{4}, H.D.B. Cooper\altaffilmark{1}, 
T.J.T. Moore\altaffilmark{3}}
\altaffiltext{1}{School of Physics and Astronomy, University of Leeds, 
Leeds LS2 9JT, UK}
\altaffiltext{2}{Max-Planck-Institut f\"ur Radioastronomie, Auf dem H\"ugel
  69, Bonn, Germany}
\altaffiltext{3}{Astrophysics Research Institute, Liverpool John Moores
  University}
\altaffiltext{4}{Leiden Observatory, Leiden University, PO Box 9513, 2300 RA
  Leiden, The Netherlands}

\label{firstpage}

\begin{abstract}
  We present the Red MSX Source (RMS) Survey, the largest statistically
  selected catalog of young massive protostars and HII regions to date.  We
  outline the construction of the catalog using mid and near infrared color
  selection, as well as the detailed follow up work at other wavelengths, and
  at higher spatial resolution in the infrared.  We show that within the
  adopted selection bounds we are more than 90\% complete for the massive
  protostellar population, with a positional accuracy of the exciting source of
  better than 2 arcseconds.  We briefly summarize some of the results that can
  be obtained from studying the properties of the objects in the catalog as a
  whole, and find  evidence that the most massive stars form: (i)
  preferentially nearer the Galactic centre than the anti-centre; (ii) in the
  most heavily reddened environments, suggestive of high accretion rates; and
  (iii) from the most massive cloud cores.
\end{abstract}


\section{Introduction}

Massive stars play a profound role in the evolution of every galaxy, but we are
still not completely sure how they form.  The origin of this problem goes back
to Kahn (1974), who realised that the radiation pressure from a young massive
star was sufficient to prevent further spherical accretion onto its surface.
The key problem is that a simple scaled-up version of the low mass star
formation model (eg Shu et al.\ 1987) gives rise to a star in which fusion
starts once it becomes more massive than $\sim$10M$\sun$, and hence significant
radiation pressure is inevitable.  In addition, massive stars are generally
observed to form in clusters, and always in high density regions, meaning the
conditions in the molecular cloud in which they form must be different from the
generic isolated low mass star.

However there are several assumptions implicit in these calculations.  First,
it ignores the role that accretion through a disc plays.  At high accretion
rates a disc will be self-shielding against ultraviolet radiation.  Secondly,
until recently, few models existed of the properties of massive stars in the
process of formation.  It was generally assumed that fusion would start once
the core of the star became massive enough but this was never actually tested.
The original conclusions of Kahn (1974) however have led to many attempts to
propose alternative models for how massive stars form.  These can be thought of
as falling into two classes: those that attempt to find a modified version of
the Shu et al.\ (1987) model, by treating disc accretion correctly following
monolithic collapse, and those that propose a completely different mode for
massive star formation.

Recent work on disc accretion has revised the previous orthodoxy that massive
stars cannot form in this fashion.  The high densities and degree of turbulence
and short timescales involved in such regions lead to accretion rates
sufficient to ensure the infalling material overcomes the radiation pressure
(eg McKee \& Tan 2003).  This led others to attempt full hydrodynamical models
including treatment of the radiation pressure.  One key aspect of all such
models is that the collimated outflows generated in the early stages provide
both an escape route for the radiation and help to sustain the turbulence in
the surrounding cloud.  Models by Krumholz et al.\ (2009, 2010), using an
approximate treatment of the radiation field, found that the role of the disc
and outflow cavity was crucial, and that the higher surface densities in
regions of massive star formation suppress break-up into clusters of smaller
stars.  Kuiper et al.\ (2010) have shown that a full radiative transfer
treatment results in their being essentially no limit to the mass of star that
can be formed, since the disc effectively self-shields against the star's
radiation pressure. More recently Kuiper \& Yorke (2013) have shown that gas,
rather than dust, opacity above the disc creates an additional shield, so that
the ``radiation pressure'' problem is reduced even further.
%

Alternative models for the formation of massive stars still have some
attractive aspects however.  Bonnell et al.\ (2004) proposed a mechanism of
competitive accretion in which a cluster forms within a turbulent molecular
cloud, with the gas being fed by dynamical processes towards the centre of the
cluster where the most massive star lies.  In this model the mass of the most
massive star is decoupled from the mass of the individual 
clump from which it forms.
Instead the mass depends on the total final cluster mass
since it is the lower mass
stars that largely ``drag'' the gas towards the centre.  Such a model naturally
leads to the mass segregation seen in actual clusters (eg Gennaro et al 2011).

The monolithic collapse model
predicts increasing accretion
rates with time to form the most massive stars, as does the competitive
accretion model.  These models do vary in other aspects 
(eg location of the most massive
accreting star in a cluster, temporal sequence of star formation throughout a
cluster).   Only a large observational sample will clearly discriminate between
them however.

Another recent theoretical development was the modelling of the internal
structure of massive stars as they are forming by Hosokawa \& Omukai (2009) and
Hosokawa et al.\ (2010).  They show that hydrogen fusion is delayed until the
star has grown to about 10$\Msolar$ at high accretion rates ($\gs10^{-4}\Msolar
$yr$^{-1}$).  Before that a combination of deuterium burning and accretion
luminosity is largely dominant.  Furthermore the star remains relatively
deficient in ultraviolet photons until accretion is almost over, since the
natural configuration for a star which has considerable mass (hence entropy)
being loaded onto it is to swell up into something resembling a cool
supergiant, a consequence that had previously also been discussed by Hoare \&
Franco (2007) in the context of the lack of HII regions around massive
protostars.  This also helps to alleviate any remaining issues with the
problem of radiation pressure in theoretical models for massive star formation.

Though much still remains to be clarified 
(e.g.\ the role of
magnetic fields), the theoretical picture is at least clearer now.
The greatest limitation therefore
is a lack of suitable observational data, especially in a
statistical sense.  
There is a clear need therefore for a comprehensive study of the young massive
star population in our Galaxy.  The ideal means of identifying young and
forming massive stars is through a combined far and mid-infrared survey, since
that is where the bulk of the stellar emission is re-radiated.  Unfortunately
existing published data covering a significant fraction of the Galactic Plane
are either of low spatial resolution (eg IRAS: cf Chan et al 1996, or Campbell
et al.\ 1989) or lack the dynamic range required to study a wide mass range (eg
GLIMPSE: Benjamin et al 2003, Churchwell et al 2009; MIPSGAL: Mizuno et
al.\ 2008, Carey et al.\ 2009).  The former means it is difficult to clearly
attribute luminosities to sources in crowded region such as the inner Galactic
Plane, and the latter makes it impossible to create a survey ranging from the
highest masses down through mid-B stars, without the former saturating on the
initial images.  Future catalogs based on Herschel (eg HiGAL, Molinari et
al.\ 2010) will alleviate both these limitations in the far-infrared and sub-mm
regimes, but such surveys cannot truly determine source properties without
additional shorter wavelength data.

Instead, as reported in Lumsden et al.\ (2002), we adopted a strategy of
combining mid-infrared data from the MSX satellite mission and ground-based
near-infrared photometry from 2MASS to characterize the properties of the
mid-infrared population of the Galactic plane.   The protostar becomes a
  bright thermal infrared source even at a relatively early stage due to the
  luminosity from contraction and accretion.  At this point dust is heated to $>200$K, which
  primarily leads to emission beyond 5$\mu$m.  Hence mid-infrared selection is
  a viable means of detecting most of the population Galaxy-wide.  We include
those highly reddened sources which are either not detected at all in the near
infrared, or else only detected at one band, in a fully self-consistent fashion
in order to ensure we include the most embedded sources.  MSX has superior
spatial resolution to IRAS, but is obviously worse than achievable with
Spitzer.  MSX is however insensitive to saturation for the objects we are
studying, unlike, say, Spitzer.   The MSX point source catalog (Egan et
  al.\ 2003) is therefore likely to be more complete, without considerable
  additional effort, than other currently existing public catalogs for the
  most massive stars.  We are primarily interested in how these stars form and
  therefore MSX is a natural choice as a starting point. The use of a single
  initial catalog is also very attractive from the point of view of
  statistical studies.  All of our final classifications are based on the
acquisition of higher spatial resolution data in the thermal infrared so the
spatial resolution of MSX is not a fundamental drawback.  The MSX data
therefore have the capacity to create a preliminary catalog from which we
draw all the young high mass objects through  extensive further follow-up
work, which has taken place over the last thirteen years.  As we noted in
Lumsden et al.\ (2002) different classes of mid-infrared bright object
segregate rather well in the combined MSX/2MASS color space.

In this paper we publish our combined catalog, including details of all the
follow-up studies carried out.  We discuss the completeness, astrometric
accuracy, classification criteria, and briefly outline some areas of science
where the catalog as a whole can help to address issues in massive star
formation.

\section{Construction of the Catalog}

\subsection{The MSX Point Source Catalog} \label{msx-intro} The Mid-course
Space Experiment (MSX) satellite mission included an astronomy experiment
(SPIRIT III) designed to acquire mid-infrared photometry of sources in the
Galactic Plane ($b<5^\circ$).  MSX had a raw resolution of 18.3$''$, a beam
size 50 times smaller than IRAS at 12 and 25$\mu$m.  MSX observed six bands
between 4 and 21$\mu$m, of which the four between 8 and 21$\mu$m are sensitive
to astronomical sources.  Full details of the mission can be found in Price et
al.\ (2001).  We used v2.3 of the MSX point source catalog (Egan et al.\
2003) as our basic input, restricting ourselves to the main galactic plane
catalog, which excludes sources seen in only a single observing pass and
those seen in multiple passes but with low significance.  The most sensitive
data, by a factor of $\sim10$, were acquired using an $8.3\mu$m filter (band A:
bandwidth 3.4$\mu$m).  As a result many objects are detected only at band A:
the vast majority of these are normal stars (see, eg, Clarke, Oudmaijer and
Lumsden\ 2005).  The point source sensitivity of band A was similar to that of
the IRAS 12$\mu$m band, at about 0.1 Jy (Egan et al.\ 2003).  Data obtained
using both the 14.7 and 21.3$\mu$m filters (bands D and E: widths 2.2 and
6.2$\mu$m respectively), although less sensitive, were also of particular value
to us since we are primarily interested in red objects.

\begin{figure}
\includegraphics[angle=-90,clip=true,width=\columnwidth]{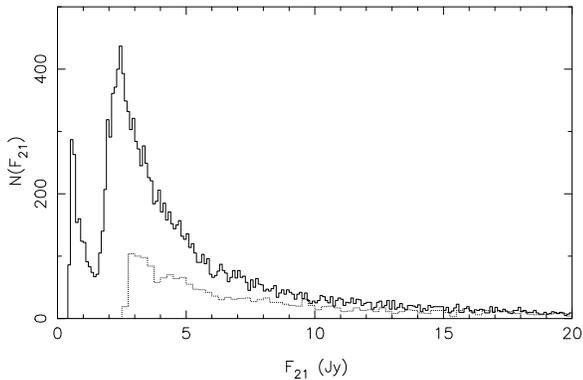}
\caption{The distribution of the 21$\mu$m fluxes for all MSX point sources
  contained in version 2.3 of the catalog in bins of 0.1Jy.  The dotted line
  shows the distribution of fluxes for all those sources confirmed by us to be
  point sources that satisfy our other selection criteria for red sources, but
  with a coarser bin width of 0.25Jy.}
\label{band_e_hist}
\end{figure}

We restricted our catalog to $10^\circ < l < 350^\circ$ in order to avoid
problems with greater source confusion, as well as kinematic distance
ambiguities, 
 near the Galactic centre.  Since we are
searching for red objects our final limiting sensitivity is that imposed by
band E.  This varies as a function of position in the galaxy (Egan et al.\
2003, Davies et al.\ 2011).  Egan et al.\ find a 90\% completeness limit of
1.5Jy for the main MSX point source catalog as a whole.  The distribution of
band E fluxes for the region of the plane we are considering is shown in Figure
\ref{band_e_hist}.  The turn-over in the number counts lies at $\sim$2.5Jy,
with reliable 95\% completeness at $\sim$2.7Jy (assuming the data follow a
power law as a function of flux).  We have adopted 2.7Jy as the limit for
our catalog.  The upturn in the data below 1.5Jy is due to detection of
sources in regions of greater sensitivity (the CB03 regions as described by
Egan et al.\ 2003).  These are limited in extent, so do not influence the
limiting sensitivity of the plane survey as a whole.

\begin{figure*}
\includegraphics[angle=0,clip=true,width=\textwidth]{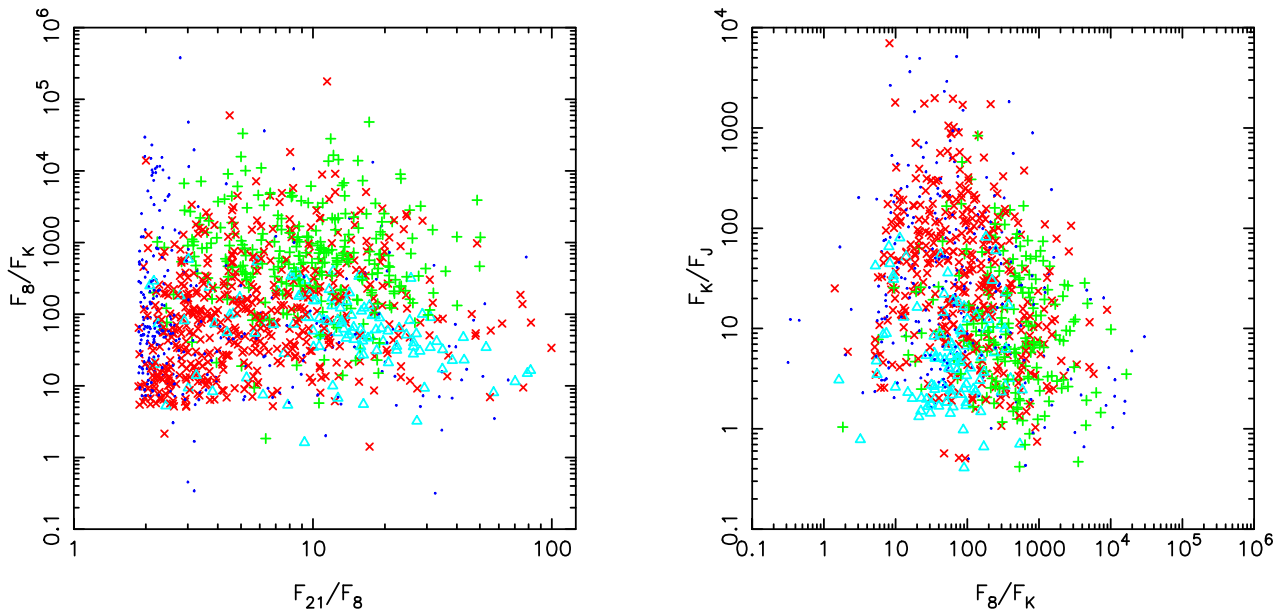}
\caption{Color-color distribution of YSOs (x, colored red in online
  version),  HII regions (+, colored green in online
  version), 
PN (triangles, colored cyan in online
  version) and evolved stars (dots, colored blue in online
  version).  The classifications are
taken from the work described in Section \ref{follow-up}.
}
\label{col-col}
\end{figure*}

Our final master list consists of those sources with quality flags of 2 or
greater in band E (which nominally corresponds to a signal-to-noise greater
than 5), and a detection in at least one of bands A and D, satisfying our
color criteria.  Where only one of A or D exists, we use the upper limit on
the other to ensure that the limit satisfies the color selection.  The final
color criteria that we apply to the MSX catalog are $F_E>2F_A$ and
$F_E>F_D$, as we discuss in Section \ref{pilot}.  We adopt conservative
boundaries in defining all the color cuts we use, whether mid- or
near-infrared, by including all objects whose error bars permit them to
possibly lie within the boundary.  The main consequence of this is that we
include a significant number of faint evolved stars in our final sample which
lie on the blue side of the boundary as is evident from the color-color plot
shown in Figure \ref{col-col}.  In total, after the MSX-based color cuts,
there are 4013 objects that satisfy our band E cut which also have detections
in bands A and D with quality $\ge2$, another 226 which are detected in bands E
and A with similar flags, but formally not detected in band D and 412 detected
in bands E and D, but not band A.  We merged all of these together into a
master list containing 4653 objects.

Hereafter in this paper we will refer to MSX bands A, D and E by their central
wavelengths as the 8, 14 and 21$\mu$m bands for simplicity.

\subsection{Extended Source Rejection} \label{eyeball} 
The MSX PSC contains many
red sources that are not genuine point sources.  The catalog parameters do
not necessarily indicate where such extended objects exist in a reliable
fashion.  A visual inspection of the MSX images revealed that $\sim1/3$ of our
candidates were more extended than 36 arcseconds.  Of course the sizes of these
sources can only be determined accurately using higher resolution data,
especially in regions where multiple real sources exist and combine to form one
MSX source.  Acquisition of higher spatial resolution mid infrared imaging 
was a key part of our survey, and we used these data in this visual rejection
of very extended sources.  In practice this employed a combination of
our own ground based mid
infrared imaging (Mottram et al 2007), Spitzer data (and specifically data from
the Spitzer Glimpse Legacy Survey is used extensively -- Churchwell et al 2009
-- as well as other archive Spitzer data where it exists) and in a few cases
where neither of these exist we recently revisited this issue by examining data
from the WISE satellite (Wright et al.\ 2010).

The complex emission present in some of these regions made it simpler to check
source sizes manually than to use an automated process.  The natural drawback
is that the final selection is prone to human error in the classification.  In
order to try and prevent the latter affecting our catalog significantly we
only excluded objects which were extended beyond approximately two MSX
beamwidths.  These sources fall into two types.

The first are genuine, internally illuminated, extended HII regions, which we
list in our catalog as ``Diffuse HII Regions''.  We are not complete for such
objects.  Sometimes very extended objects are split into more than one source
by the MSX point source finding algorithm.  In such cases we have kept only a
single final source.  We note however that the astrometry of such sources is
particularly poor and we have not attempted to improve on the original MSX
co-ordinates.  We have identified 620 sources classified in this way.  These
``Diffuse HII Regions'' are retained in the final catalog, though in general
we have not studied them further.

The second are actually rejected from the master list since they appear to be
either extended background or filamentary emission around much larger
structures, or artifacts due to the presence of nearby bright sources.  There
are 1501 such sources.  Emission of this kind naturally belongs to the larger
source and is not a point source on its own. Such sources tend to cluster in
particular around sites of known extended star formation (eg the Carina
nebula).  In these regions it appears the MSX point source detection algorithm
is not reliable.  For the same reason we are probably missing genuine point
sources in such areas, especially closer to the brighter central regions.  In
Davies et al.\ (2011) we attempted to quantify this affect by randomly placing
point sources into the original MSX image tiles and then recovering them using
a similar algorithm to the MSX PSC.  The result is that our final catalog
completeness is actually position dependent on the sky.  Davies et
al.\ factored this into their statistical analysis of the catalog.

Finally visual inspection of high resolution mid infrared data allows us to
identify the main counterpart of the MSX source correctly, and hence improve on
the MSX astrometry, as well as addressing the issue of source multiplicity.  We
will return to this issue in Section \ref{astrometry}.  Perhaps most crucially,
this check has also allowed us to identify compact clusters of exciting
sources.  Hereafter we only consider those 2539 objects that pass the
requirements that any candidate has at least one red compacct source present in
the MSX beam.

\subsection{Near Infrared Counterparts}\label{2mass}
The only available suitable all-Plane
 survey in the near infrared is 2MASS (Skrutskie et
al.\ 2006).  We utilised the data from both the Point Source Catalog (2MASS
PSC) and Extended Source Catalog (2MASS XSC).  The 2MASS detectors are
limited at bright magnitudes by saturation, but fluxes in the PSC are derived
from fits to the wings of the point spread function reducing that limitation to
a handful of very bright stars.  At faint magnitudes, the 2MASS PSC is however
close to confusion limited in many areas of the Galactic Plane.

We have also now added deeper infrared imaging data to our database where
available.  These data come from two surveys, the UKIRT Infrared Deep Sky
Survey (UKIDSS) Galactic Plane Survey (GPS: Lucas et al.\ 2008) and the Vista
Variables in the Via Lactea Survey (VVV: Minitti et al.\ 2010), both of which
are deeper and have better spatial resolution than 2MASS.  UKIDSS uses the
UKIRT Wide Field Camera (Casali et al, 2007) on the UK Infrared Telescope.  The
UKIDSS project as a whole is defined in Lawrence et al (2007).  The pipeline
processing and science archive for UKIDSS/VVV are described in Hodgkin et al
(2012), and Hambly et al (2008).  For the data used in this paper, a similar
procedure was in use for VVV (Lewis, Irwin and Bunclark 2010).  The photometric
system of UKIDSS is described in Hewett et al (2006), and the calibration is
described in Hodgkin et al. (2009).

\begin{figure*}
\includegraphics[angle=0,clip=true,width=6in]{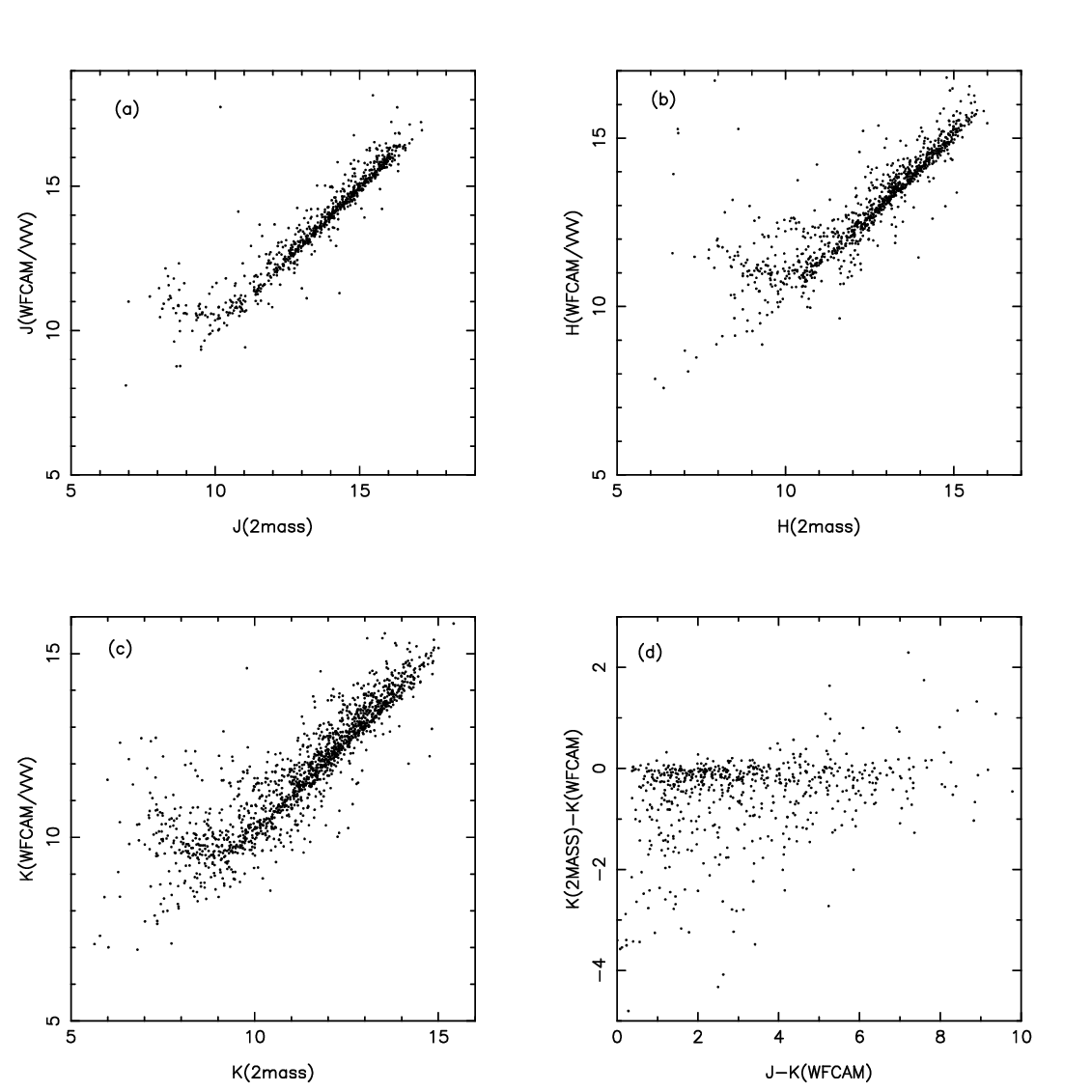}
\caption{Comparison of 2mass and UKIDSS or VVV magnitudes, for (a) $J$,
(b) $H$, (c) $K$.  The final panel, (d) shows the color term involved in
comparing UKIDSS and 2MASS magnitudes at $K$.
}
\label{2mass-wfcamvvv}
\end{figure*}

There are relatively minor differences in the $J$ and $H$ band photometry
between 2MASS and UKIDSS/VISTA, but a more substantial one at $K$ where both
2MASS and VISTA use a similar $K_s$ filter, and UKIDSS a more traditional $K$
filter which additionally permits the transmission of light in the
$\sim2.3-2.4\mu$m range.  The color terms to translate these data back onto
the 2MASS system are known for relatively unreddened stars (eg Hodgkin et al
2009), 
but rely on having accurate $J$ 2MASS data which is not the case for many of
our very red sources.  We have considered the color term for all unsaturated
$K$ band sources with UKIDSS and the results are shown in Figure
\ref{2mass-wfcamvvv}.  The majority of our sources are consistent with only a
small color term of $\sim0.2(J-K)$ being required to place the UKIDSS data
fully on the 2MASS system.  Since the errors in the magnitudes in these regions
tend to be as large as this small correction we have ignored it in what
follows.

We also plot the global comparison of the 2MASS data with the combined
UKIDSS/VVV data in Figure \ref{2mass-wfcamvvv}.  The 2MASS data will always be
preferable for very bright objects since the UKIDSS and VISTA cameras saturate
near 10th magnitude in all bands as can be seen in the Figure.  There also
exists a substantial group of objects where the data differ by more than the
nominal errors on either set.  These largely comprise objects which are blended
with either other stars or nebular contamination in 2MASS, though there are a
subset which are actually fainter in the 2MASS survey.  Some of these are due
to saturation effects in the UKIDSS/VVV data, and others are moderately
extended HII regions in the near infrared, so the catalogued magnitudes depend
largely on the measurement method.  In practice, we do not include near
infrared data for clearly very extended objects such as most HII regions in our
analysis. There are also occasional examples of stars that would appear to be
extremely variable (e.g.\ the low mass YSO G338.5459+02.1175 has genuinely
become one hundred times fainter between the 2MASS and VVV observations).
These exceptions also help to explain the presence of unusual objects in the
comparison of 2MASS and UKIDSS photometry as a function of UKIDSS color.
Wherever possible in the final catalog we have adopted UKIDSS/VVV data, where
they exist, for sources with 2MASS magnitudes fainter than the saturation
limits, and 2MASS data otherwise.

The actual near infrared counterpart selected is based on the co-ordinates
derived from the mid-infrared astrometry (see Section \ref{astrometry}).  We
selected the best match based on visual comparison of the mid and near-infrared
data.  Some objects have no direct counterpart, since they are obscured even at
$K$, though extended emission such as reflection nebulosity may be present.
All these objects are guaranteed to be sufficiently red to satisfy the color
criteria described in Section \ref{pilot}.  Of course this matching still
leaves the possibility that some of our sources are not visible in the near
infrared and that we have picked a neighbouring near infrared source which has
a chance alignment at the resolution of the data with the real target. This may
be especially true in massive star forming regions where dense clusters of
associated stars are present (e.g Carpenter et al.\ 1993; Hodapp 1994) and in
some cases the true MYSO may be completely obscured.  Such dense clusters
however appear to be rare in our sample, and we are therefore confident that
the number of incorrectly attributed near infrared fluxes remains small.

 We can partly quantify these effects.  We identified all 2MASS sources
  within a 5 arcsecond radius of the 2539 sources that passed the checks
  discussed in the previous section.  These include many sources, such as most
  HII regions, where there is no clear counterpart near the centre.  The number
  of false associations can be estimated by extrapolating back the number seen
  at larger separations, which are all presumed to be chance coincidence, to
  those within 1 arcsecond.  The net result is that perhaps 10\% should be seen
  by chance within 1 arcsecond.  This drops to about 2\% within 0.5 arcseconds.
  Our typical astrometric accuracy from the mid-infrared lies in this range for
  point sources.  We have excluded any near infrared counterpart that lies
  outside this range as being a likely chance coincidence.  The actual ratios
  are remarkably similar when we consider the deeper UKIDSS data, with 15\%
  expected by chance within 1 arcseconds, and 3\% within 0.5 arcseconds.  The
  UKIDSS/VVV astrometry is accurate enough that we can match to within the
  smaller radius.  Finally, we should note that every counterpart has been
  visually inspected to ensure that the identification is sensible.  Where it
  appears to be a chance alignment (eg a very blue star with a very red thermal
  infrared source) we have excluded the cross-match.  Overall we expect this
  leaves us with no more than 5\% incorrectly identified counterparts.


\subsection{Astrometry}\label{astrometry}
The general astrometry of the MSX point source catalog is discussed by Egan
et al.\ (2003).  Improvements were made to the earlier version of the MSX PSC
which meant positions of isolated point sources are reasonably accurate.
Issues still arise near very bright sources, as discussed in Lumsden et al.\
(2002).  However the main concern with the astrometry lies in crowded regions.
We have checked the basic reliability for point sources in several ways.  The
most uniform comparison, in the sense that it covers all of our RMS sources,
was with the all-sky WISE survey, which matches the longer wavelength
range of MSX but with a better spatial resolution (12 arcseconds at 25$\mu$m,
and 6 arcseconds at all the shorter wavelengths).  WISE suffers from saturation
for bright sources and the fact that the published catalog is described as a
source finding catalog rather than a point source catalog (Cutri et al.\
2012).  Nonetheless, we have cross-correlated the original MSX and WISE
catalogs for point sources that are fainter than the absolute WISE saturation
limit (about $F_{25}=330$Jy).  We first considered the positions of evolved
stars (which tend to be isolated): 99.5\% of the sources have positions that
agree within 5 arcseconds, and 80\% agree within 2 arcseconds.  For all sources
(excluding diffuse HII regions and those rejected during visual inspection) the
equivalent figures are 9.5 and 3.5 arcseconds.  Excluding those HII regions
that are resolved by WISE but not strongly by MSX reduces these values by
0.5--1 arcseconds.  Most of the difference between the values for the evolved
stars and other sources arises due to source multiplicity in crowded regions --
all such regions are associated with star formation.

It is clear therefore that we required improved astrometry to accurately
identify counterparts.  This is not possible to do at a long enough wavelength
that we are sampling near the peak of the spectral energy distribution, but it
is possible to use thermal infrared images to identify counterparts, from which
shorter wavelength data can then be used to improve the astrometry in most
cases.  The main surveys used to improve the accuracy of the astrometry in this
way were the Spitzer GLIMPSE, 2MASS, radio images taken during the construction
of the catalog, UKIDSS and VISTA VVV
surveys, and finally, if other options proved unsuitable, the WISE survey or
Spitzer MIPS images.

As the discussion above demonstrates, there is generally more than a single
source that is bright at thermal infrared wavelengths in star forming regions,
as well as significant extended emission.  Our aim was to identify all
significant compact sources present within, or very close to, the MSX beam that
can contribute to the total 21$\mu$m luminosity.  These ``multiple sources''
are separated in our online catalog, with the astrometry measured as for
other objects.  The luminosity for each individual source is estimated using
the highest spatial resolution thermal infrared data.  These correction factors
are all listed in the online version of our catalog.  There are a few
exceptions to this process.  Extended HII regions can have rather irregular
morphologies, and it is difficult to determine whether there is only one region
present.  We also ignore any sources in the field that are ``blue'' since they
will contribute relatively little to the far infrared luminosity.  Finally, we
have not separated sources where more than one evolved star is present, as
these are not of primary interest to us.  We have tried to separate all sources
where a YSO is present with either another YSO or an HII region however.  The
catalog currently contains 216 original MSX candidates that have been
separated in this fashion.

The process of choosing the best astrometry varies slightly with source type.
Where the target shows a clear point source in GLIMPSE or equivalent we have
adopted that position.  Where Spitzer data are unavailable we have matched the
thermal infrared information we do have (either our own or based on WISE) to
select potential near infrared counterparts if these exist.  We note that we
only use such near infrared data where we can locate the source with reasonable
accuracy from the thermal infrared -- near infrared data on their own are
unlikely to be entirely accurate as we may instead pick reflection nebulosity
as the centre where a very obscured source is hidden from our line of sight.
We only use WISE or MIPS data for point source astrometry when these conditions
are not met.  For HII regions or PN that are compact, we adopt the same
procedure, unless they are detected in the radio when we adopt the radio
position.  Weak, irregular large HII regions pose a particular challenge as
they have no clearly defined centre.  In such circumstances we adopted the flux
weighted centroid from GLIMPSE, WISE or MIPS data if that is feasible, or else
mark their centre using the position of the apparent exciting star(s).  The
accuracy of such results may only be good to 2--3 arcseconds when comparing
with results at different wavebands, but is generally better than MSX.

Currently in the online database, 2781 objects have had their astrometry
refined, 1104 using GLIMPSE or other Spitzer IRAC data, 942 using 2MASS, 115
from UKIDSS or VVV data, 398 using radio positions, 203 using WISE.  The
remainder are a combination of MIPS, two from ground based thermal
infrared images where no other alternative was suitable, and 18 which still
retain their original MSX positions.

\begin{figure}
\includegraphics[angle=-90,clip=true,width=\columnwidth]{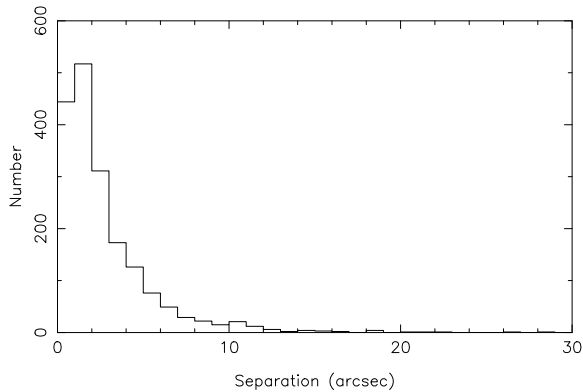}
\caption{The distribution of the offsets between the catalog MSX position
  and that derived from data with higher spatial resolution.
}\label{final-original-astrometry}
\end{figure}

Figure \ref{final-original-astrometry} shows the offset from the initial MSX
position to the final adopted positions for this subset of data.  The offsets
seen that are greater than 15 arcseconds mostly correspond to sources which we
have split into two or more counterparts, with the exception of the well known
HII region G333.6032--00.2184, where even from the MSX image the original
position is clearly inaccurate.  There are however sources of all kinds that
have not been split, where the astrometry differs in the range 10--15
arcseconds, including those where the counterpart is highly reliable since it
is drawn from the GLIMPSE data.  In all such cases a visual inspection of the
available GLIMPSE images reveals a complex background with significant extended
emission as well as point sources present.  Overall however the results show
that 80\% of objects fall within 4.5 arcseconds and 99\% within 13 arcseconds.
We stress however that the final astrometry, since it uses much higher spatial
resolution data, is good to better than an arcsecond in almost all cases.

 The final catalog of potential candidates was only drawn up after this
  astrometric identification of individual counterparts was completed.
  Inevitably this meant that we had partial ``working versions'' of the
  catalog along the way, which wrongly excluded some objects, generally on
  the grounds of having appeared extended in MSX images alone, whereas higher
  resolution data suggests a more point-like source is present.  These objects
  tend to lack the same depth of coverage in the follow-up observations
  outlined in Section \ref{follow-up}, but have been added back into the final
  catalog with an appropriate classification.

\subsection{The Pilot Survey and Final Color Selection Criteria}\label{pilot}
Our initial multi-color selection criteria for picking massive protostars from
the MSX catalog was published in Lumsden et al.\ (2002).  That paper relied
on object classifications available in the literature.  This could have led to
a bias, since few of the literature classifications were the results of
systematic surveys.  In order to determine the reliability of our source
selection using the MSX and 2MASS catalogs we therefore initially carried out
a small pilot survey of about 100 sources, all of which lay in the outer galaxy
in order to reduce confusion.  The objects were selected on a single color cut
of $F_{21}/F_{8}>1.5$, with MSX quality flags set at 4 in both bands.  We did
not impose a flux or luminosity limit, though in practice the quality threshold
implies $F_{21}$ greater than about 5Jy.  This selection encompasses the
previously published color selection but includes bluer objects, and does not
rely on matching with a near infrared counterpart.


We obtained our own near and mid-infrared imaging data using the UK Infrared
Telescope, as well as spectroscopy of a smaller subsample from the same
facility.  The spectroscopic data will be published separately in Cooper et
al.\ (2013), together with all the other spectra obtained at UKIRT as part of
the main RMS survey, and both mid and near-infrared images are available on our
database. Here we are only interested in the overview that these data gave on
our color selection.  Many of the sources within the pilot survey had already
been studied in detail, so that we could draw upon literature classifications
to aid in this process, with the spectroscopy also informing our final
identifications.  The full classification procedure that these data helped
inform is discussed in detail in Setion \ref{follow-up}.



The net result of this survey therefore is that our initial supposition from
Lumsden et al.\ (2002) that we can use a combination of MSX mid infrared data
and suitable near infrared data was correct.  The final near and mid-infrared
color boundaries we adopted on completion of this survey are the same as those
published in Lumsden et al., namely: $F_{21} > 2F_8$, $F_{21}>F_{14}$,
$F_{14}>F_8$, $F_8>5F_K$ and $F_K>2F_J$.  The depth of the initial 2MASS 
catalog
guarantees that if we do not detect these objects at $K$ then the available
upper limit does satisfy the $F_8>5F_K$ constraint.  Therefore we include all
these non-detections in our final color selected sample.  The same is true for
detections at $K$ but not $J$ in general as well.

The mid-infrared boundaries essentially just set the constraint that we expect
all YSOs to have rising infrared continua in this wavelength range.  The main
advantage of adding near infrared data is to help discriminate against objects
which have detached dust shells.  In that case the mid infrared data may look
like a YSO, since it arises from the warm dust in the shell, but the near
infrared generally reveals the central sources with rather bluer colors than
any embedded YSO.  The online database also contains data and classifications
for objects satisfying only the mid-infrared color cuts, but which fail the
near infrared cuts.  We have not subsequently studied such objects in detail as
many are evolved stars.

\subsection{Follow-up Observations and Object Classification}\label{follow-up}
The goal of the RMS survey was to accurately characterize all of the MSX point
sources which passed the near and mid-infrared color selection.  The basic
observational dataset
that exists for the majority of these are as follows: higher resolution mid
infrared imaging than available from MSX alone (either ground based 10$\mu$m
observations, e.g.\ Mottram et al.\ 2007, or
publically available Spitzer data as noted previously); higher resolution, and
deeper, near infrared data than available from 2mass as noted in Section
\ref{2mass}; longer wavelength far infrared and submm data in order to
constrain the spectral energy distribution, all taken from existing public
archives (see Mottram et al 2010, 2011a); ${}^{13}$CO $J=1-0$ or $J=2-1$ mm
line observations in order to determine kinematic distances (Urquhart et al
2007a, 2008); radio images to study the HII region and planetary nebulae
populations (either our own imaging data -- Urquhart et al 2007b, 2009
-- or the images present in the CORNISH radio survey -- Hoare et al.\ 2012,
Purcell et al.\ 2013); and finally, for those with appropriate characteristics,
$H+K$ band spectroscopy of the identified near infrared counterpart in the case
where that counterpart is sufficiently bright in the $K$ band (Cooper et al.\
2013).  In all cases the published results should always be seen as a
snapshot of the contents of the full database at that time.  When newer data
have become available we have incorporated these in to the online version of
the catalog which is the definitive source.  The database pools all of this
information into a single collection for each source, including images at all
wavelengths where we have access to such data.  We plan to keep the database
updated as new information appears for the foreseeable future.

The final source classification following these observations is decided
individually for every source.  The full online version of the database
contains our reasoning for each of these.  The basic idea though is
as follows:

A source that is extended at $8-10\mu$m with no obvious point like core is
likely to be an HII region or planetary nebula.  This is because the dust
morphology tends to follow the ionised gas in such regions, rather than being
concentrated around the central star(s) (eg Hoare, Roche \& Glencross 1991).
Young stellar objects, where the central core is hidden even at 10$\mu$m, can
also be slightly extended at the highest resolutions, since we then only see
the thermal emission from cavity walls (eg de Wit et
al.\ 2010).  There is generally a clear distinction between these sources
however.  

Any source detected in the radio at a level at $\sim5$GHz much above 10mJy must
be an HII region or planetary nebula (see also Figure \ref{radio-plot} which we
have used to refine this process).  There are examples of known massive YSOs
with detected radio emission (eg Guzm{\'a}n et al.\ (2010) found an example from
the RMS survey), but all lie below 10mJy (see also the discussion in Hoare \&
Franco 2007).  It does however illustrate that it is important to be careful in
the use of this criterion alone for weak radio sources.  These need to be
considered according to the expected radio flux as a function of luminosity as
shown in Figure \ref{radio-plot}.

We classed as stellar any source that has a point-like mid-infrared core
(including those with diffuse emission around a point-like core), unless it had
significant radio emission.  The latter are likely to be ultracompact HII
regions (or very young planetary nebulae), if they exceed the flux limit noted
above.

The previous criteria serve to identify  extended HII regions and PN, and
radio bright point-like HII regions and PN.  The remaining sources
were all point-like, and are likely to be either YSOs or evolved stars.  Some
may be very compact PN or HII regions, but only if they fall below our radio
detection threshold ($\sim1$mJy in most cases) -- Figure \ref{radio-plot}
shows we are only seriously incomplete in the radio below about
$L_{bol}\sim10^4$\Lsolar (most of the non-detections above this luminosity are
for objects that are over-resolved by our radio surveys).

Objects that have unambiguous far infrared counterparts (whether from IRAS or
MIPSGAL) which reveal a peak in the spectral energy distribution at 25$\mu$m
are likely to be evolved.  This is due to the presence of detached dust shells
around most such sources.  This class includes most known PN, and all but the
most embedded evolved stars.

Any source that is undetected in $^{13}$CO $J=1-0$ is highly likely to be
evolved since molecular gas is commonplace in star formation regions.  The
converse, that no evolved source shows CO emission, is not however true.  Where
we have CO emission we derived kinematic distances in the fashion described in,
e.g., Urquhart et al.\ (2008), using the rotation curve from Reid et al.\
(2009).  For sources within the solar circle the kinematic distance ambiguity
also has to be solved.  A discussion of this can be found in Urquhart et al.\
(2012), and we note that we also adopted solutions from Green et al.\ (2011)
where their methanol maser sources lie in the same fields and with the same
$V_{LSR}$ as our RMS sources.  We derived luminosities from the spectral energy
distribution (Mottram et al 2011a) using these derived distances.

Young massive stars generally form in clusters.  They also tend to show
evidence for reflection nebulosity, outflows, dust lanes and other features
typical of star formation regions.  These are rarer for the main types
of evolved star which we are sensitive to.  Therefore, if a source 
shows extended emission or clustering in either the near or mid-infrared,
it is likely to be young, and if
it does not it is likely to be evolved.

We also use our near infrared spectroscopy in determining the final class.  For
example, there are some sources which reveal an isolated stellar point source
in the near infrared, but with considerable nebulosity in the field.  Spectra
obtained of a small sample of these reveal about an equal split between evolved
stars (presumably seen through an obscuring veil that fills the field) and
young stars.  The remaining objects in this sub-group
for which we have not as yet acquired spectra are
listed in the catalog as young/old sources.

PN and HII regions can both also be point like, and below our radio limit, as
noted above.  Most such sources have had near infrared spectra acquired.
Ionised gas emitting under standard case B recombination
leads to a higher equivalent width of the Br$\gamma$ line of
hydrogen (Cooper et al.\ 2013) allowing a relatively clear discrimination from 
the generally weaker Br$\gamma$ seen in YSOs.

Another criterion that can be used to distinguish young and old sources is the
presence of maser emission.  Methanol masers are characteristic of young
massive stars (eg.\ Green et al.\ 2009).  The presence of double peaked OH
maser emission is the characteristic classification technique for OH/IR stars
(eg.\ Eder, Lewis and Terzian, 1988), whilst single-peaked or more irregular
line profiles can arise in star formation regions.

At the end of this process we weigh all these factors and decide a final class
based on them.  For most objects this process is relatively unambiguous.  Some
however have classes that remain uncertain (especially those where we lack 
portions of the follow-up data).  

The final catalog lists 11 categories of objects classified on the basis of
these criteria.  These comprise five evolved star groups (generic evolved
stars, PN, proto-planetary nebulae, OH/IR stars and carbon stars), four that
are demonstrably young (YSOs, HII regions, objects that appear to have
characteristics of both of these which we class as HII/YSO and diffuse HII
regions), the young/old sources mentioned above,  and finally a group of 77
objects that are classified as other (another generic catch-all for objects of
known type that do not fall into any other category).

Some of these categories are not necessarily complete.  For example, we list
known carbon stars separately.  Many of the objects we list as evolved stars
may also be carbon stars, but we lack the data to tell this.  Therefore the
complete list of all carbon stars in our sample will contain those classified
as such together with some subset of the larger group classified simply as
evolved stars.  The same is true for OH/IR stars.  
We have focused much more on
the young stars, where our classification is rather more rigorous.

The final classifications were used in producing the color-color diagrams
shown in Figure \ref{col-col}.  We have only plotted objects where data exists
to form both colors for clarity.  There are obviously also many objects with
only upper limits at $J$, $K$ and even 8$\mu$m that are not shown here.  We
have also plotted only four broad classes of objects on these diagrams.  The
evolved stars include only the ``catch-all'' evolved star class in the
catalog, plus the carbon and OH/IR stars.  There is a clear sense that the
majority of the evolved stars defined in this way tend to be rather blue in the
thermal infrared.  There are a smaller set of genuinely highly reddened
evolved stars that do not follow the standard OH/IR or carbon star tracks in 
Figure \ref{col-col} (cf the discussion in Lumsden et al.\ 2002).
The nature of these evolved stars is not clear, though it appears to
exclude all the known carbon stars.  A few known OH/IR stars do lie within this
genuinely red population.

The other three classes of objects shown are those HII regions that are compact
enough to be ``unresolved'' by MSX, PN (all detected PN are relatively compact)
and all YSOs, regardless of luminosity.  The near and mid-infrared colors do a
remarkably good job of separating the PN from the HII regions, consistent with
the picture outlined previously that PN illuminate optically thin dust shells,
whereas young HII regions all live in optically thick environments.  Finally
the YSOs fill most of the color space.  This clearly demonstrates the value of
detailed follow-up observations, since it would be impossible to color-select
our YSO population otherwise without significant contamination by evolved
stars,
planetary nebulae 
and HII regions.  Our spectroscopy of the massive YSOs suggests this wide
color dispersion is probably an evolutionary effect, in the sense that the
YSOs become bluer and the dust disperses as they age (Cooper et al.\ 2013).

\subsection{Completeness}\label{completeness}
In Lumsden et al.\ (2002) we estimated that we should detect a B0 star anywhere
in the galaxy from the sensitivity of the original MSX point source catalog.
In Davies et al.\ (2011) we refined these estimates by determining whether we
could recover false sources injected into MSX images, and hence defined a
completeness as a function of position in the galaxy.  Here we consider again
whether those estimates were correct by comparing the original MSX data with
that from the much more recent WISE mission and also by comparing our final
catalog with other published samples of massive YSOs or UCHII regions.

\begin{figure}
\includegraphics[angle=0,clip=true,width=\columnwidth]{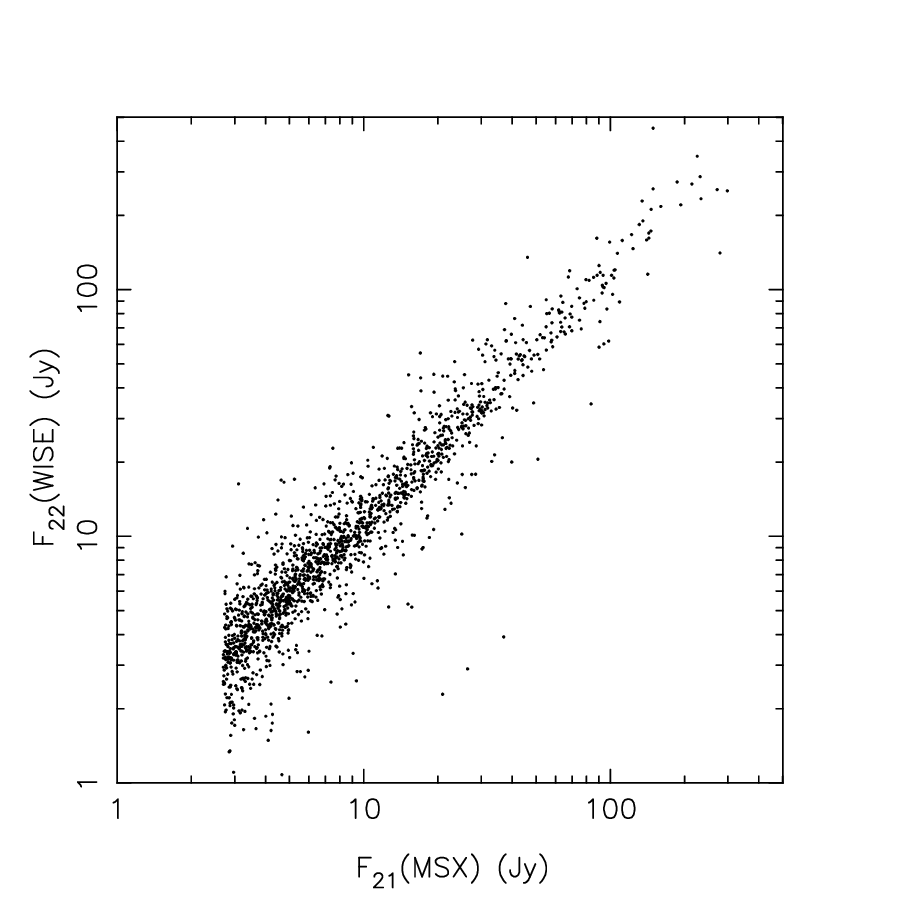}
\caption{The comparison of the 21$\mu$m fluxes for all MSX point sources,
  with quality flag $\ge3$ at 21$\mu$m, with WISE 22$\mu$m fluxes for sources 
  that can be defined as pointlike, with signal-to-noise $>5$ and mostly
  unsaturated. }
\label{f21-f22-plot}
\end{figure}

The WISE data have better spatial resolution at 10$\mu$m than MSX by a factor
of 3, and are considerably deeper in limiting flux.  WISE does have a
relatively faint saturation limit however, and the published catalog is not a
complete {\em source} catalog so much as a flux-detection catalog.  We
first compared the 21$\mu$m fluxes from MSX for all our good sources with the
22$\mu$m sources found in WISE.  We only considered sources with MSX quality
flag of 3 or 4 at 21$\mu$m, and limited the WISE counterparts to those that
have less than 25\% of saturated pixels, are relatively point-like (see below)
and with the signal-to-noise ratio in the 22$\mu$m band $>5$.  Figure
\ref{f21-f22-plot} shows the result for 1658 matched sources.  There is clearly
a good correlation, with a fair degree of scatter due mostly to extended
sources.  This suggests the MSX data are indeed a good basis for our catalog.

We can also use the WISE data to derive a ``RMS-like'' sample with the same $l$
and $b$ boundaries to test the completeness of our catalog.  We used data
with signal-to-noise of at least 5 in all bands, since WISE is much more
sensitive than MSX.  We further restricted ourselves to compact sources by
requiring the $\chi^2$ profile fitting values from the WISE catalog to be
less than 15.  Figure \ref{f21-f22-plot} suggests we restrict ourselves to
$F_{22}>2.8$Jy for a fair comparison.  Finally, equivalent color cuts for our
RMS sources are approximately $F_{22}>2F_{12}$, $F_{22}>2F_{4.6}$,
$F_{4.6}>1.3F_{3.4}$ and $F_{22}>10F_K$.

There were 228 objects satisfying these criteria which were not in our RMS
catalog.  Of these 156 had MSX data.  On inspection all but 10\% of these are
genuine point sources.  Those with counterparts show that 97 have MSX
$F_{21}<2.7$Jy, excluding them from our RMS selection, 24 were in the MSX
Singleton (ie single detection)
list which we did not use, and 101 had MSX colors that fail our
color selection (generally inverted in $F_{14}/F_8$ or $F_{14}/F_{22}$
ratios).  Objects such as these, which are marginally outside our initial MSX
selection criteria, are unlikely to effect our conclusions.  For example, if we
consider just the flux limit alone, a similar number of sources have been
included in our sample which are in reality fainter than our stated flux limit
(as easily seen in Figure \ref{f21-f22-plot}).  Of the remaining 72 WISE
objects which are not detected by MSX, a visual inspection reveals that only 13
are definitely real, with another 7 possibly so.  The others are a mixture of
extended objects, regions around saturated objects, and a few very confused
regions.

One group of WISE detections that we excluded from RMS 
are of greater interest.  We excluded
objects from our original selection that were detected only at 21$\mu$m, as our
experience with MSX suggested single band detections, other than at the deeper
8$\mu$m band, were unreliable.  There are 27 objects detected only at 21$\mu$m
by MSX which also satisfy the WISE color cuts above however.  These all form
an unusual category of object, where the $F_{22}/F_{12}$ ratio is larger than
other sources in RMS, but the $F_{4.6}/F_{3.4}$ ratio is fairly normal.  The
12$\mu$m images suggest these objects all sit in the core of dark clouds, where
the silicate absorption feature is dominating the spectral energy distribution.

Overall therefore we can say that WISE finds additional sources that at most
amount to about 50 objects, or about 2--3\% of our total final red MSX sample.
We therefore conclude that our initial completeness estimates are reasonable
for the color selection we adopted.

We can also compare our catalog with other published lists of young massive
stars.  A complete catalog for ultracompact HII regions exists from the
recently completed CORNISH blind radio survey (Purcell et al.\ 2013).  A
high-reliability compact HII region catalog (Urquhart et al.\ 2013b) with
clear ATLASGAL counterparts (Schuller et al.\ 2009) has recently been produced.
This indentified 207 compact and ultracompact HII regions that lie within the
union of the boundaries of CORNISH and ATLASGAL.  We recover 160 of these in
our final list of good RMS sources.  A further 18 have other MSX counterparts
within 10 arcseconds, with 10 with 21$\mu$m flux below our limit, seven with
brighter 8$\mu$m emission than 14$\mu$m (suggestive of strong PAH emission
commonly seen around HII regions, but excluded from our color selection)
and one only detected by MSX at 21$\mu$m (which as noted above we excluded from
consideration for RMS).  Notably, many of these 18 also have stronger 12$\mu$m
emission than either 8 or 14$\mu$m, indicative of strong 12.8$\mu$m [NeII]
emission from their nebulae.  Sources such as these are often extended in MSX,
and would have been classified as Diffuse HII regions by us.  Inspection of the
Spitzer Glimpse and MIPSGAL images for the sources that have no MSX
counterparts reveal an equal combination of extended sources that would have
also been classified by us as Diffuse HII regions, and faint sources in complex
backgrounds.   Overall we conclude that the use of our color constraints
misses about 10\% of the total HII region population within our size limits.  
We used the modelling of Davies et al.\ (2011) to estimate the fraction
missing due to the size constraint itself.  Less than 5\% 
of ultracompact HII regions are missing anywhere in the galaxy, consistent
with our comparison with the CORNISH catalogue.  The larger compact HII regions
are less well represented, with more than 50\% missing, though presumably some
fraction of these in reality make up our Diffuse HII region catgeory.  
Therefore our overall size
constraint
is strictly only useful in terms of detecting ultracompact HII regions.

We can also compare with the earlier, IRAS-based, Chan, Henning \& Schreyer
(1996) and Sridharan et al.\ (2002) catalogs of high-mass protostellar
candidates.  For the Chan et al.\ catalog, if we excluded those outside our
survey area, only 14 of an initial 215 are not present in our catalog.  We
examined these in detail.  All but one are either fields where the IRAS
position falls between two well separated regions of star formation (which are
in our catalog), or are genuine objects detected in MSX where $F_8>F_{14}$.
Such sources also appear to be multiple HII regions, but largely unresolved by
MSX.  The colors are a reflection of the properties of the individual sources.
Virtually none are point sources at the resolution of WISE or Spitzer.  The
exception to this is IRAS 18079--1756, which has no counterpart in the MSX
data.

The Sridharan et al.\ catalog should be a ``cleaner'' sample of massive YSOs,
given the greater constraints they place on object selection.  In particular,
they specifically select against objects with known radio emission in single
dish surveys.  We find a match for 40 of the 69 sources within 20 arcseconds.
Examination of the remaining 29 sources shows that 21 have MSX counterparts
within the same radius.  The overwhelming majority of these MSX counterparts
have inverted spectra, consistent with strong PAH and [NeII] emission effecting
the flux at 8 and 12$\mu$m.  The same sources also appear significantly
extended in WISE images.  Overall these appear to be classical extended HII
regions, {\em ie} exactly the sources that Sridharan et al.\ tried to select
{\em against}.  Indeed many show evidence of radio emission as well (from
CORNISH but also MAGPIS -- specifically Helfland et al.\ 2006). The remaining 8
sources appear to have very large scale diffuse emission at 22$\mu$m, with
multiple possible exciting sources contained within.  We would classify all but
4 objects as Diffuse HII regions.  Only one source, IRAS 05553+1631, is a
genuine compact source not present in the MSX catalog.  Again, within our
stated selection criteria there is relatively little evidence of large
scale incompleteness in our results for compact infrared bright sources.

Another possible comparison, which may better target the YSO population, is
provided by the methanol multibeam survey (MMB: Green et al.\ 2009).  Methanol
masers are known to appear only in regions of high mass star formation.  The
whole of the southern portion of the methanol multibeam survey is now published
(see Green et al.\ 2012).  Only 202 of the 554 methanol maser sources that lie
within $187^\circ<l<350^\circ$ appear in our initial RMS candidate list (with a
search radius of 20 arcseconds).  Using a tighter 4 arcsecond radius, and
considering only matches with our compact sources (ie not the diffuse HII
regions), reduces this total to 141, of which 93 are YSOs and 48 HII regions.
However, MMB fails to detect emission from 348 of our YSOs, and 506 of our HII
regions, within the current published MMB region.  Clearly MMB alone is not a
sufficient means of detecting all massive young stars.  

In order to determine whether the MSX ``dark'' MMB sources are truly missing we
also looked for them in the WISE catalog.  This recovers 
 529 of the 554 MMB sources with the same 20
arcsecond search radius.  Approximately half of
these have $F_{22}<2.7$Jy.  The tighter 4 arcsecond search radius reduces this
overlap to 355 of 554.  These values are consistent with the finding of
Gallaway et al.\ (2013) who did a similar analysis using the Glimpse point
source catalog.  Gallaway et al.\ also show that the MMB sources without
counterparts in our catalog are on average fainter at 8$\mu$m, consistent
with our findings here (they also find they are often bluer than our RMS
selection criteria). The luminosity distribution of the matches with RMS is
also informative in this analysis, since it shows that 43 of the 141 have
$L\le10^4$\Lsolar.  The objects that are genuinely ``dark'' are clearly of
interest, since they may represent a heavily embedded phase that we
are not fully sensitive to, but these only form about 10\% of the MMB
population (Gallaway et al).  The others are consistent with a population of
lower luminosity sources, or much more distant sources with luminosity
$\sim10^4$\Lsolar (ie just below our completeness limit).  Urquhart et al.\
(2013a) present a more detailed analysis of the counterparts of these maser
sources.  Their findings are
consistent with the simple analysis presented here.

Overall therefore the evidence suggests that the RMS survey is greater than
95\% complete within its color and flux cuts.   This is true regardless of
  which alternative survey we compare the RMS survey against.  In addition, we
  have also quantified the incompleteness due to our actual selection criteria.
  The main contributor to missing ``young'' sources comes from unusual colours,
  and from the failure to recover very red sources.  We conclude that the RMS
  survey recovers greater than 90\% of all compact young massive sources with
  mid-infrared emission.  

\section{The Catalog}\label{catalog}
The RMS catalog is continuously being updated as new information becomes
available.  In particular we are still refining distances when new information
appears.  The current catalog uses the Reid et al.\ (2009) model that
accounts for the known parallax distances (e.g.\ Rygl et al.\ 2010).  Where
individual sources have themselves been studied using maser parallaxes, we also
adopt the parallax distance.  The initial kinematical information is all
provided in the database to allow users to adopt the Galactic rotation model of
their choice.

In addition, we are extending the wavelength coverage of the spectral energy
distribution and adding data with better spatial resolution when it becomes
available.  The on-line database includes images of our fields as well as
spectra, all downloadable as FITS files.  In addition a search function makes
it possible to query the catalog directly, allowing users to select
sub-samples approrpiate to themselves.  We are always interested in adding
functionality that will make the RMS database more useful, and welcome
suggestions for extensions.

The only definitive version of the catalog is therefore online,
at http://rms.leeds.ac.uk/

The version presented here provides a subset of the catalog information.
Table 1 lists RMS name, position, source type, a flag indicating whether the
source passed the near infrared color cuts (all sources by definition pass the
MSX cuts), $v_{LSR}$, kinematic near and far distance, and adopted distance
(which if different, indicates a non-kinematic origin for the distance such as
parallax), Galactocentric radius, luminosity and IRAS counterpart if one
exists.  The color cut is defined by identifying {\em blue} objects with the
simple prescription $F_K/F_J<(2-\delta(F_K/F_J)$ or
$F_8/F_K<(5-\delta(F_8/F_K)$, where the $\delta$ represent the errors in the
derived ratio.  These represent objects which must be too blue to fall in
our sample.  We use the best near infrared data available for each source in
calculating these ratios.

\begin{table*}
\setlength{\tabcolsep}{2pt}
\begin{tabular}{lllllrrrrrrl}
\hline
  \multicolumn{1}{c}{RMS Name} &
  \multicolumn{1}{c}{RA} &
  \multicolumn{1}{c}{Dec} &
  \multicolumn{1}{c}{Type} &
  \multicolumn{1}{c}{Color} &
  \multicolumn{1}{c}{$v_{LSR}$} &
  \multicolumn{1}{c}{Near} &
  \multicolumn{1}{c}{Far} &
  \multicolumn{1}{c}{Adopted} &
  \multicolumn{1}{c}{$R_{gc}$} &
  \multicolumn{1}{c}{Lbol} &
  \multicolumn{1}{c}{IRAS} \\
  \multicolumn{1}{c}{} &
  \multicolumn{1}{c}{} &
  \multicolumn{1}{c}{} &
  \multicolumn{1}{c}{} &
  \multicolumn{1}{c}{Cut?} &
  \multicolumn{1}{c}{(kms$^{-1}$)} &
  \multicolumn{1}{c}{(kpc)} &
  \multicolumn{1}{c}{(kpc)} &
  \multicolumn{1}{c}{(kpc)} &
  \multicolumn{1}{c}{(kpc)} &
  \multicolumn{1}{c}{(\Lsolar)} &
  \multicolumn{1}{c}{} \\
\hline
  G010.0997+00.7396 & 18:05:13.10 & -19:50:34.7 & PN & N &  &  &  &  &   &  &  \\
  G010.1089+00.3716 & 18:06:36.30 & -20:00:51.4 & PN & Y &  &  &  &  &   &  & 18036-2001\\
  G010.1481-00.4260 & 18:09:39.55 & -20:21:59.6 & Evolved star & N &  &  &  &  &   &  &  \\
  G010.3204-00.2616 & 18:09:23.29 & -20:08:06.9 & HII region & Y & 32.1 & 3.8 & 12.8 &  &   &  & 18064-2008\\
  G010.3207-00.2329 & 18:09:17.25 & -20:07:18.2 & HII region & N & 32.5 & 3.8 & 12.8 &  &   &  &  \\
  G010.3208-00.1570A & 18:09:00.36 & -20:05:08.6 & HII region & Y & 12.0 & 2.0 & 14.6 & 3.5 & 4.95 & 84500 &  \\
  G010.3208-00.1570B & 18:09:01.48 & -20:05:08.0 & YSO & Y & 12.0 & 2.0 & 14.6 & 3.5 & 4.95 & 41620 &  \\
  G010.3844+02.2128 & 18:00:22.67 & -18:52:09.7 & YSO & Y & 5.5 & 1.1 & 15.5 & 1.1 & 7.35 & 1180 & 17574-1851\\
  G010.3930+00.5389 & 18:06:34.27 & -19:41:05.3 & PN & Y &  &  &  &  &   &  & 18036-1941\\
  G010.4413+00.0101 & 18:08:37.94 & -19:53:59.0 & HII region & Y & 66.4 & 5.3 & 11.2 & 11.2 & 3.3 & 40020 & 18056-1954\\
\hline
\end{tabular}
\caption{The full RMS catalog.  This table  is published in its
  entirety in 
  the electronic edition. A portion is shown here for guidance regarding
  its form and content. }
\end{table*}

Table 2 presents the subset of massive protostars which pass all color cuts,
and additionally have $L_{bol}>20000$\Lsolar.  This table gives position,
adopted distance, luminosity and commonly used ``Other Name''.  These names are
only given for sources with more than ten citations on SIMBAD for counterparts
within 20 arcseconds of the central RMS source.   We present all the 
objects we classify purely as YSO, as well as those which have clear
characteristics of HII regions (eg strong radio emission), but where the
central exciting star still retains characteristics of a YSO, such as spectral
evidence for a disk (see, eg, Cooper et al.\ 2013).  Only about half of our
most luminous protostars are sufficiently well studied to have a common ``Other
Name''.  This clearly illustrates the additional benefits provided by the RMS
survey for the study of luminous protostars.

\section{The Properties of Young Massive Stars}

More detailed analysis of some of the bulk properties are presented in our
other papers already.  For example, we have modelled the young massive star
population of the Galaxy, and our observational data then rule out models in
which the accretion rate decreases with time (Davies et al.\ 2011).  We also
find a maximum luminosity for a massive YSO of about $2\times10^5$\Lsolar, that
is consistent with the Davies et al.\ and Hosokawa et al.\ (2010) models
(Mottram et al.\ 2011b).  We can derive lifetimes and luminosity functions for
both the massive YSO and compact HII region phases (Mottram et al.\ 2011b).
We have also mapped the distribution
of the young massive star population in the Galaxy and shown how well it
correlates with the spiral arm structure (Urquhart et al.\ 2011a).

Here we consider some of the properties that the catalog as a whole reveals
about the young massive stellar population of the Milky Way that utilize the
breadth of the data from our follow-up observations.

\subsection{The Anomalous Radio Luminosity of Young B-stars}
\begin{figure*}
\includegraphics[angle=0,clip=true,width=\textwidth]{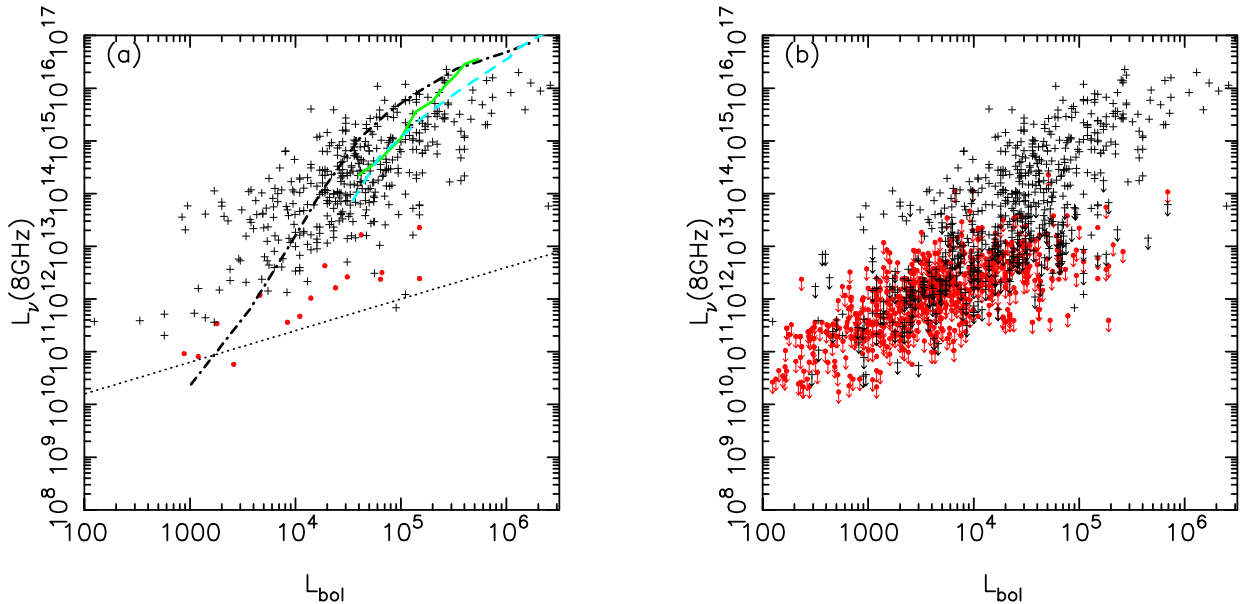}

\caption{Radio luminosity versus bolometric luminosity for all of the HII
  regions and YSOs in the catalog, (a) showing only the detections, and (b)
  showing all data including limits. The HII regions are crosses, the YSOs dots
  (colored red in the online version).  Model radio luminosities are also
  shown in (a) after the fashion outlined in Lumsden et al.\ (2003).  The
  dot-dash black line shows the results for a single star using stellar models
  from Lanz \& Hubeny (2007), the dashed (blue in online version) line shows
  the results for a single star using stellar models from Pauldrach, Hoffmann
  \& Lennon (2001), and the solid (green in online version) line the Pauldrach
  et al.\ models but for a cluster. The dotted line shows the extrapolation of
  jet luminosity as derived for low mass stars by Anglada (1995), as well as of
  the known emission from high mass stars (whether jet or wind) as given
in Hoare \& Franco (2007).}
\label{radio-plot}
\end{figure*}

In Figure \ref{radio-plot} we show the radio versus bolometric luminosity.  A
clear envelope to the detected HII region emission can be seen in the figure,
which must represent the maximum optically thin free-free emission.
We can compare this to the predictions from OB stellar atmosphere models.  We
have plotted three model sequences here.  The first two use the WM-basic models
of Pauldrach, Hoffmann \& Lennon (2001), in the fashion outlined in Lumsden et
al.\ (2003), for a single central exciting star and for a cluster of such stars
following a standard initial mass function.  The last model sequence uses the
same data as applied in the modelling of Davies et al.\ (2011), with the 
stellar models coming from Martins et al.\  (2005) and Lanz \& Hubeny (2007).
In all cases we convert the model $N_{LyC}$, the number of Lyman continuum 
photons, to a radio flux using equation 4 from Kurtz, Churchwell and Wood
(1994):
\begin{align}  S_\nu({\rm Jy})  = &  1.32\times10^{-49}  \epsilon a 
N_{LyC} \nonumber \\ & \times 
\left(\frac{\nu}{{\rm GHz}}\right)^{-0.1} \left(\frac{T_e}{{\rm
      K}}\right)^{0.5} \left(\frac{d}{{\rm kpc}}\right)^{-2} \end{align}
 where
$\epsilon=1$ is the fraction of the Lyman continuum ionising flux that actually
ionises the gas (as opposed to exciting dust, or simply leaking away),
$a\sim0.98$, and we adopt an electron temperature $T_e=8000$K.  

The detected HII regions with $L\ls5\times10^4$\Lsolar\ all tend to show more
emission than the models.  Since this is a luminosity-luminosity plot, the
issue cannot lie wholly with inaccurate distance.  Errors in the fluxes used in
estimating the bolometric luminosity are likely to move points to the left if
anything, since we still rely in part on large-beam IRAS data for objects
outside the MIPSGAL region.  Objects can be fainter than the envelope, since
optical depth, leakage of ionising photons from the nebula and over-resolved
interferometric data all act to give less flux than expected.  Nothing however
acts to make an object brighter in the radio than anticipated.  The fact the
envelope is so well defined, even at low luminosity, strongly suggests this is a
real effect.  

Significant excess Lyman continuum flux above model predictions has
been observed directly in EUV observations of the nearby stars
$\epsilon$ and $\beta$ CMa (Cassinelli et al.\ 1995, 1996). These are
B2 II and B1 II--III spectral types and hence are not on the main
sequence and thus perhaps not direct analogues of the young B
stars ionizing UCHII regions. However, if these young stars have not
completely contracted onto the main sequence they may not be so
dissimilar.  The timescales for the contraction onto the main sequence
after the termination of accretion are of order 10$^{5}$ years for
early B stars (Figure 2 in Davies et al. 2011), which is similar to the
age of the UCHII regions themselves.

The physical reasons behind the order of magnitude excess Lyman continuum flux
are still being sought. An infrared excess in the B2
II stars showed that the outer layers of the atmosphere were warmer than the
models predict. Krticka, Korcakova and Kubat (2005) discuss the effects of
Doppler and frictional heating of stellar winds that are significant in the
cooler B stars. The hot, X-ray dominated winds in some late O star main
sequence stars may also be of relevance here (Huenemoerder et al. 2012).

Finally, Figure \ref{radio-plot}(b) clearly shows that most of the detected
massive YSOs lie below the HII region detections.  The radio emission from
these objects is not nebular.  The stars simply lack ultraviolet flux (as is
also evident from the near infrared spectra presented by Cooper et al.\ 2013).
They are therefore ``radio weak'' compared to the HII regions with the same
luminosity.  This deficit of ultraviolet photons is in agreement with models
for massive YSOs which predict the star swells and cools as it accretes (eg
Hosokawa \& Omukai 2009).  Two possible alternatives then exist for the source
of the radio emission.  We may be seeing emission from an ionised jet (see, eg,
Figure 5 in Anglada 1995, which we have fitted together with the known high
mass radio detections in YSOs from Hoare \& Franco (2007), 
to derive a relationship through the luminosity
regime in Figure \ref{radio-plot}).  This perhaps lies at the base of any
larger scale molecular outflow.  Alternatively, we may be seeing an
equatorially flattened wind (eg Hoare 2006).  The relative paucity of actual
detections in our existing radio data, as well as their relatively modest
spatial resolution, makes it impossible to distinguish between these cases.

\subsection{Virialised Motions in Massive Cores?}
As part of the follow-up observations for the RMS survey we have acquired a
significant quantity of new mm CO line data in order to 
derive kinematic distances to our sources.
The line intensities were not fully calibrated, but the velocities and line
widths are.  

\begin{figure*}
\includegraphics[clip=true,width=\textwidth,angle=0]{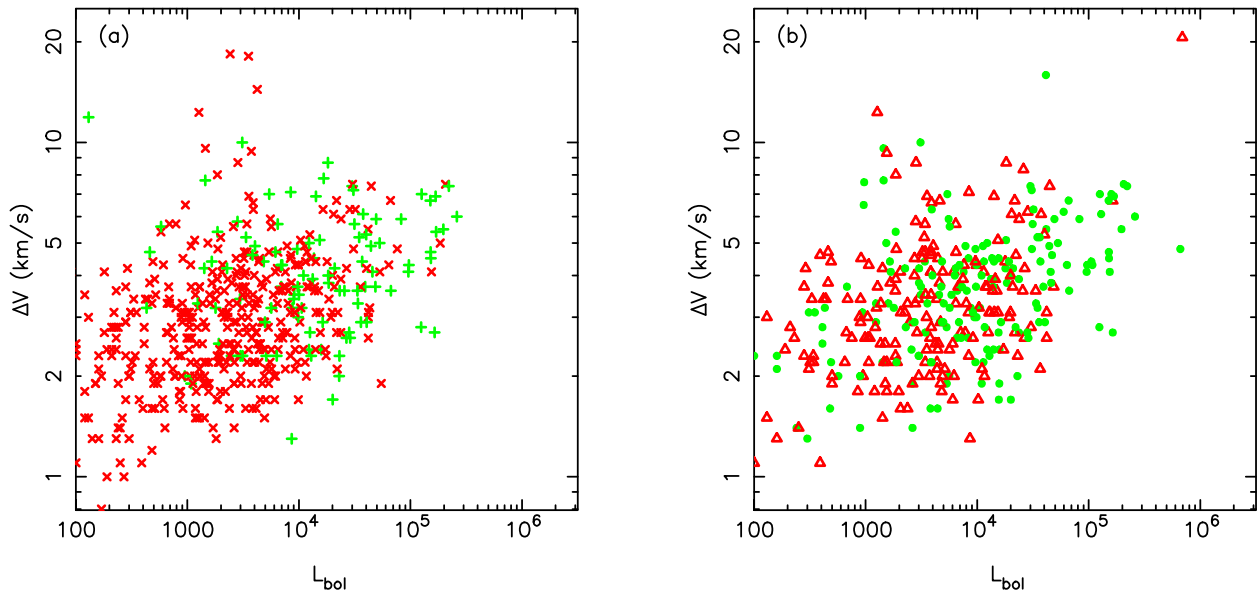}

\caption{Bolometric luminosity versus linewidth of our observed $^{13}$CO data.
The sample is limited to those objects with distances nearer than 5kpc for
comparison with the results of Figure \ref{lumcol}.  
The data for YSOs (x, colored red in the
online version) and HII regions (+, colored green in the online version)
are shown in (a), and 
the same set of data but split by color in (b), 
with the points (green in the online version)
having $F_{21}/F_8>10$ and the triangles (red in the online version)
$4<F_{21}/F_8<10$.  
  }
\label{linewidth-lum}
\end{figure*}

In Figure \ref{linewidth-lum} we plot the relationship for HII regions and YSOs
between the full width at half maximum of the $^{13}$CO emission and the
bolometric luminosity.  We have curtailed the sample to those with distances
within 5kpc, and limited the HII regions to those objects which appear
point-like in the WISE data (see Section \ref{completeness} for details).  The
left hand plot shows where HII regions and YSOs lie separately, the right hand
plot shows all HII regions and YSOs, but instead shows where objects fall as a
function of the $F_{21}/F_8$ ratio, where we have split the data into two
around $F_{21}/F_8=10$.  The more luminous objects in Figure
\ref{linewidth-lum} have the broader lines.  The formal Pearson correlation
coefficient between $\log(\Delta v)$ and $\log L$ is  $\sim0.43$ 
for 523 objects.
Urquhart et al.\ (2011b) found a similar relationship using the width of the
NH$_3$ line for an overlapping but not identical sample of YSOs and HII regions
drawn from RMS.

 The correlation holds separately for YSOs, with a correlation coefficient
  of 0.41 for 422 YSOs.  By comparison the HII region sample on its own shows
  no significant evidence for a correlation (correlation coefficient of 
0.10 for 101 HII regions), rising to 0.19 if the restriction on 
``point-like'' HII regions is removed).  It is possible that the
the bulk expansion motions seen in the gas around HII regions destroys any
correlation there.  Not all of our YSOs show evidence for an outflow (Maud et
al.\ in preparation), which suggests that the good correlation seen for the
YSOs is actually a fundamental property of the natal molecular cloud rather
than due to the energetics of the central source. 

Urquhart et al.\ suggested that this relationship was due to the correlation
between linewidth and the mass of the natal clump (ie it is essentially a
virial relation, eg Larson 1981), and that seems the most natural explanation
here too.  Potentially, this also explains the weak color segregation present
in $F_{21}/F_8$ as well, since we would expect the objects that will become
more massive to have higher accretion rates, and hence have redder colors in
the thermal infrared.  High accretion rates can only be sustained if the core
itself is more massive, naturally leading to redder objects having somewhat
enhanced linewidths.

\subsection{Galactocentric Radial Distribution of RMS Sources}
\begin{figure}
\includegraphics[angle=0,clip=true,width=\columnwidth]{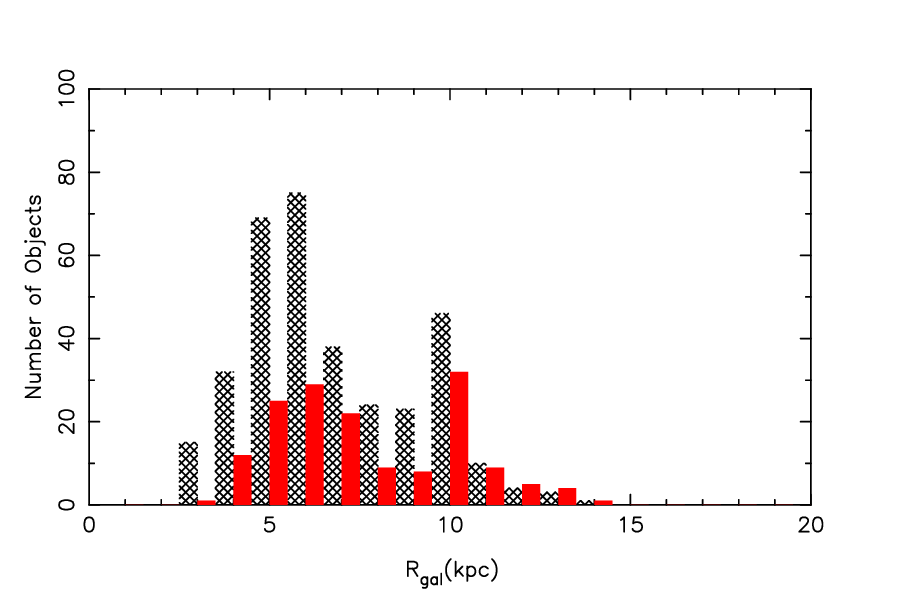}
\caption{Distribution of galactocentric radius for HII regions (grey hatched
  bars) and YSOs (solid, colored red in online version).  Only sources with
  bolometric luminosity greater than $10^4\Lsolar$ and distance from us of less
  than 10kpc are included.  }
\label{rg-plot}
\end{figure}

The distribution of the RMS sources in the Galaxy has previously been
considered by Urquhart et al.\ (2011a), but we are now in a position to
accurately study how this varies with the type of source.  In Figure
\ref{rg-plot} we show the distribution of both HII regions and YSOs as a
function of Galactocentric radius.  We have restricted the sample to those
objects with $L_{bol}>10000$\Lsolar, and distance from us of less than 10kpc to
match the rough completeness limit of the catalog.  A Mann-Whitney U-test
suggests these differ in the sense that the YSOs lie at larger radii at the
99\% confidence level.  This could partly be a selection effect.  It is easier
to see bright radio emitting HII regions in the considerable background near
the Galactic centre than it is a YSO.  There is a lack of HII regions in the
outer galaxy compared to the YSOs however, and this cannot be a selection
effect.  In part this also appears to be a luminosity effect.  If we compare
the YSOs against the HII regions with $L_{bol}<100000$\Lsolar, the confidence
level drops to 95\%.  This value was chosen as the approximate upper limit of
YSO luminosity.  By contrast the confidence level that the samples are
different is $>99$\% when we compare the higher luminosity HII regions with the
YSOs.  There is only one HII region with luminosity above this threshold beyond
a Galactocentric radius of 10kpc -- the other 62 all lie inwards.

The most obvious explanation if this result is correct is that high mass cores
are more likely in the crowded central regions.  Lepine et al.\ (2011) have
found a similar ``step-change'' in numbers of open clusters and the metallicity
gradient at a similar Galactocentric distance of 8.5kpc.  They explain it as
being due to a gap in the dense gas at the co-rotation radius, that isolate
properties in the inner Galactic Plane from those outside the co-rotation
radius.  Their explanation requires smaller gas flows outside the co-rotation
radius, which will lead to less collisions between cloud cores.  It seems
plausible that the most massive star formation regions arise through such
collisions (eg.\ van Loo, Falle and Hartquist 2007), giving a natural
explanation for the result we see.

\subsection{Color-Luminosity Relationships}
\begin{figure*}
\includegraphics[angle=0,clip=true,width=\textwidth,angle=0]{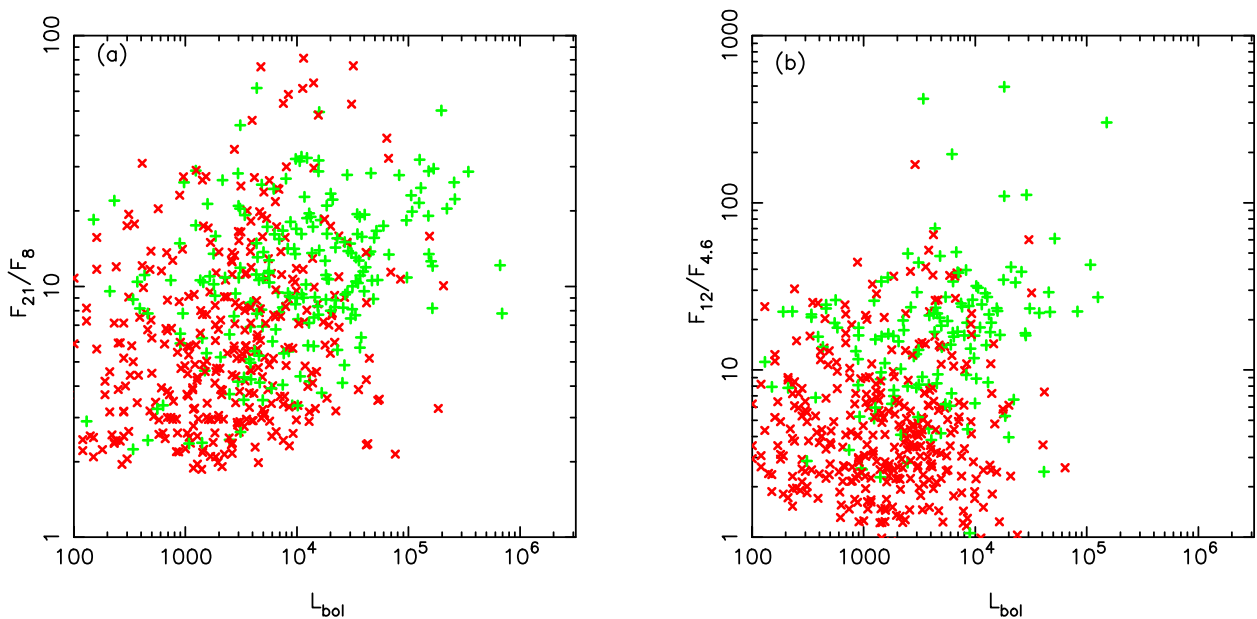}

\caption{Bolometric luminosity versus mid-infrared color.  The left hand plot
  shows data from the MSX satellite, the right data from WISE.
The sample is limited to those objects with distances nearer than 5kpc.
The   YSOs are shown as  x (colored red in online version) and HII regions as
+ (colored green in online version).  There is a clear
sense in which the HII regions are redder than the YSOs on average.
  }
\label{lumcol}
\end{figure*}

Finally we consider how the colors of the YSOs and HII regions vary as a
function of luminosity.  Figure \ref{lumcol} shows data using the MSX
$F_{21}/F_8$ ratio as well as the ratio from WISE bands 3 and 2 (12 and
4.6$\mu$m).  The latter is a useful test since the beam size for these
wavelengths is a factor of three smaller than for the MSX data.  Again we have
curtailed the sample to those sources which appear point-like in WISE at
12$\mu$m, and which lie within 5kpc.  The latter is important here since we see
more HII regions to greater distances due to their greater luminosity on
average.  The line of sight extinction, and hence to some extent the thermal
infrared color, increase with distance creating a bias.

The HII regions, on average, are redder than the YSOs in both ratios.  This
points towards a picture in which HII regions are more embedded. The only other
factor that can play a part is geometry.  This comes into play in two ways.
Orientation is important in determining the colors of a YSO, since a more
edge-on accretion disc source will suffer greater extinction than one we see
pole-on.  However the data shown should randomly sample all inclinations.  This
may give rise to the large scatter in color seen, but cannot explain the
difference between the HII regions and the YSOs.  The second factor is the
extent of the HII region.  More evolved HII regions, as they grow larger,
naturally support a greater volume of Ly$\alpha$ heated dust.  This should
shift the colors bluewards, since there is more emission away from the centre
of the natal core.  Again part of the scatter could be due to this effect.
Since this factor should make HII regions bluer however, it cannot explain why
HII regions are actually redder than YSOs.  The colors, like the virial
estimates, again suggest more massive cores, which have greater line of sight
reddening towards the centre of the core, give rise to higher
luminosity sources.

\section{Summary}

The final catalog from the Red MSX Source Survey presents a comprehensive
view of the infrared bright phase of the massive protostellar population of our
galaxy.  It is complete for objects as faint as a typical B0 star at the
distance of the Galactic centre and for more luminous sources across the whole
galaxy.  We currently have almost 700 YSOs and HII regions with confirmed
luminosities above this threshold.  In addition we have identified many more
YSOs that can be characterised as embedded Herbig Be stars that lie below this
luminosity, providing a valuable, though incomplete, resource for the study of
young intermediate mass stars as well.  The combination of a bright saturation
limit for the initial MSX catalog together with the coverage of the whole
Galactic Plane out to $|b|<5^\circ$ makes it the ideal catalog for these
infrared bright sources.  Comparison with both older lower spatial resolution
catalogs of massive young stellar objects derived from IRAS, and more recent
higher spatial resolution candidates from both Spitzer and WISE, as well as
high resolution radio studies, suggests that our catalog is highly reliable
and complete for the massive young stellar object class.  It is likely to
remain the best source for this infrared bright phase for some time, given the
relatively low saturation limits of the IRAC and MIPS instruments from Spitzer,
and to a lesser extent for the detectors on WISE.  This compromises the ability
of those missions to study the most massive star formation regions and most
luminous sources.

The full RMS catalog has also been used by other groups for aspects of
research into massive star formation.  Guzman et al.\ (2010, 2012) have studied
the radio properties of objects they classify as ``radio weak'', in a bid to
detect hypercompact HII regions, jets and stellar winds.  Thompson et al.\
(2012) examined the incidence of triggering using RMS sources around
bubbles found in Spitzer data.  Moore et al.\ (2012) studied the correlation of
available gas reservoirs with our MYSOs in order to determine how star
formation efficiency varies as a function of location within the spiral arms.
Davies et al.\ (2012) used the distances we have derived to our YSOs and HII
regions to help pin down the distances to massive star clusters.  In addition,
other groups have adopted similar color selection criteria to those present in
Lumsden et al.\ (2002) in order to study heavily obscured evolved stars (eg.\
Ortiz et al 2005).  Our catalog contains a large number of highly obscured
evolved stars likely to be of interest to those interested in topics such as
post-AGB evolution.

We have also shown that the wealth of data we have acquired can themselves
identify both new properties (the excess UV flux from young intermediate mass
stars), as well as provide indications as to how massive stars form.  The
correlation between luminosity and apparent virial mass suggests that these
dynamical processes are essentially responsible for setting the initial
condition rather than the ongoing formation.  This tends to favour models of
monolithic collapse, though competitive accretion may play an important role at
an earlier phase when the clumps themselves are merging, and possibly also in
feeding residual mass to the most massive star present one it has formed.  The
split between inner and outer galaxy may also signify the greater role that
triggering and spiral density waves can play in denser environments.

Finally we note that the RMS catalog is already the basis for numerous further
studies in massive star formation, such as the properties of outflows.  When
combined with appropriate catalogs of the initial phases of massive star
evolution from other indicators (eg.\  Caswell et al.\ 2010,
Contreras et al.\ 2013, Molinari et al.\ 2010), a
complete census, and understanding of the evolution,
of the massive star population from initial core through to
main sequence will be possible.  This should provide firm answers to the 
remaining questions as to how massive stars form.

\section{Acknowledgments}

We would like to thank the referee for their comments.  This work was supported
by the STFC.  SLL also acknowledges the support of PPARC through the award of
an Advanced Research Fellowship during the early stages of this work.  This
publication makes use of data products from the Two Micron All Sky Survey,
which is a joint project of the University of Massachusetts and the Infrared
Processing and Analysis Center/California Institute of Technology, funded by
the National Aeronautics and Space Administration and the National Science
Foundation.  This research has made use of the SIMBAD database, operated at
CDS, Strasbourg, France.  This work is based in part on observations made with
the Spitzer Space Telescope, obtained from the NASA/ IPAC Infrared Science
Archive, both of which are operated by the Jet Propulsion Laboratory,
California Institute of Technology under a contract with the National
Aeronautics and Space Administration.  This publication makes use of data
products from the Wide-field Infrared Survey Explorer, which is a joint project
of the University of California, Los Angeles, and the Jet Propulsion
Laboratory/California Institute of Technology, funded by the National
Aeronautics and Space Administration.

\parindent=0pt

\vspace*{3mm}

\section*{References}
\begin{refs}
\mnref{Anglada G., 1995, RMxAC, 1, 67 }
\mnref{Benjamin R.A., et al., 2003, PASP, 115, 953 }
\mnref{Bonnell, I.A., Vine, S.G., Bate, M.R., 2004, \mnras, 349, 735 }
\mnref{Campbell B., Persson S.E., Matthews K., 1989, AJ, 98, 643 }
\mnref{Carey S.J., et al., 2009, PASP, 121, 76 }
\mnref{Carpenter J.M., Snell R.L., Schloerb F.P., Skrutskie M.F.,
        1993, ApJ, 407, 657}
\mnref{Casali M., et al., 2007, A\&A, 467, 777 }
\mnref{Cassinelli, J.P., et al., 1995, ApJ, 438, 932 }
\mnref{Cassinelli, J.P., et al., 1996, ApJ, 460, 949 }
\mnref{Caswell, J.L., Fuller, 
G.A., Green, J.A., et al.\ 2010, \mnras, 404, 1029 }
\mnref{Chan S.J., Henning T., Schreyer K., 1996, A\&AS, 115, 285 }
\mnref{Churchwell E., et al., 2009, PASP, 121, 213 }
\mnref{Clarke A.J., Oudmaijer R.D., Lumsden S.L., 2005, MNRAS, 363, 1111 }
\mnref{Contreras, Y., et al., 2013, A\&A, 549, A45 }
\mnref{Cooper H.D.B., et al., 2013, MNRAS, 430, 1125 }
\mnref{Cutri R.~M., et al., 2012, ``Explanatory Supplement to the WISE All-Sky
  Data Release Products'' }
\mnref{Davies B., Hoare M.G., Lumsden S.L., Hosokawa T., Oudmaijer R.D., 
Urquhart J.S., Mottram J.C., Stead J., 2011, MNRAS, 416, 972 }
\mnref{Davies, B., et al., 2012, MNRAS, 419, 1871 }
\mnref{de Wit W.J., Hoare M.G., Oudmaijer R.D., Lumsden S.L., 2010, A\&A,
  515, A45} 
\mnref{Eder J., Lewis B.M., Terzian Y., 1988, ApJS, 66, 183 }
\mnref{Egan M.P., Price S.D., Kraemer K.E., Mizuno D.R., Carey S.J., 
Wright C.O., Engelke C.W., Cohen M., Gugliotti G. M., 2003, 
The Midcourse Space Experiment Point
Source Catalog Version 2.3 Explanatory Guide,
    Air Force Research Laboratory Technical Report AFRL-VS-TR-2003-1589}
\mnref{Gallaway M., et al., 2013, MNRAS, 626 }
\mnref{Gennaro M., Brandner W., Stolte A., Henning T., 2011, MNRAS, 412, 2469 }
\mnref{Green, J.A., Caswell, 
J.L., Fuller, G.A., et al.\ 2009, \mnras, 392, 783}
\mnref{Green, J.A.,  McClure-Griffiths, N.M.\ 2011, \mnras, 417, 2500 }
\mnref{Green, J.A., Caswell, 
J.L., Fuller, G.A., et al.\ 2012, \mnras, 420, 3108 }
\mnref{Guzm{\'a}n A.E., Garay G., Brooks K.J., 2010, ApJ, 725, 734 }
\mnref{Guzm{\'a}n, A.E., Garay, G., Brooks, K.J., Voronkov, M.A.\ 
2012, \apj, 753, 51 }
\mnref{Hambly N.C., et al., 2008, MNRAS, 384, 637 }
\mnref{Helfand, D.J., Becker, R.H., White, R.L., Fallon, A., Tuttle, S. 
  2006, AJ, 131, 2525 }
\mnref{Hewett P.C., Warren S.J., Leggett S.K., Hodgkin S.T., 2006, MNRAS,
  367, 454 }
\mnref{Hoare, M.G., 2006, ApJ, 649, 856 }
\mnref{Hoare M.G., Roche P.F., Glencross W.M., 1991, MNRAS, 251, 584 }
\mnref{Hoare, M.G., Franco, J., 2007, In: ``Diffuse Matter from Star 
Forming Regions   to Active Galaxies - A Volume Honouring John Dyson'', 
Edited by T.W. Hartquist,
  J.M. Pittard, and S.A.E.G. Falle. Series: Astrophysics and Space Science
  Proceedings. Springer Dordrecht, 2007, p.61}
\mnref{Hoare M.G., et al., 2012, PASP, 124, 939 }
\mnref{Hodapp K.-W., 1994, ApJS, 94, 615}
\mnref{Hodgkin S.T., Irwin M.J., Hewett P.C., 
Warren S.J., 2009, MNRAS, 394, 675 }
\mnref{Hodgkin S., Irwin M., Lewis J., 
Gonzalez-Solares E., Yolda{\c s} A.K., 2012, Star Clusters in the Era of Large
Surveys, Astrophysics and Space Science Proceedings, Springer-Verlag,
 Berlin Heidelberg,  p39 }
\mnref{Hosokawa T., Omukai K., 2009, ApJ, 691, 823 }
\mnref{Hosokawa T., Yorke H.W., Omukai K., 2010, ApJ, 721, 478 }
\mnref{Huenemoerder, D.P., Oskinova, L.M., Ignace, 
R., Waldron, W.L., Todt, H., Hamaguchi, K., Kitamoto, S., 2012, ApJ, 756, L34 }
\mnref{Kahn F.D., 1974, A\&A, 37, 149}
\mnref{Krumholz, M.R., Klein, R.I., McKee, C.F., Offner, S.S.R., 
   \& Cunningham, A.J., 2009, Science, 323, 754 }
\mnref{Krumholz, M.R., Cunningham, A.J., Klein, R.I., 
McKee, C.F., 2010, ApJ, 713, 1120}
\mnref{Krticka, J., Korcakova, D., Kubat J., 2005, PAICz, 93, 29 }
\mnref{Kuiper, R., Klahr, H., Beuther, H., Henning, T., 2010, ApJ, 722, 1556 }
\mnref{Kuiper, R., Yorke, H.W., 2013, ApJ, 763, 104}
\mnref{Kurtz S., Churchwell E., Wood D.O.S., 1994, ApJS, 91, 659 }
\mnref{Lawrence A., et al., 2007, MNRAS, 379, 1599 }
\mnref{Lanz T., Hubeny I., 2007, ApJS, 169, 83 }
\mnref{Larson R.B., 1981, MNRAS, 194, 809 }
\mnref{L{\'e}pine J.R.D., et al., 2011, MNRAS, 417, 698 }
\mnref{Lewis J.R., Irwin M., Bunclark P., 2010, ASPC, 434, 91 }
\mnref{Lucas P.W., et al., 2008, MNRAS, 391, 136 }
\mnref{Lumsden S.L., Hoare M.G., Oudmaijer R.D., Richards D., 2002, MNRAS, 
336, 621 }
\mnref{Lumsden S.L., Puxley P.J., Hoare M.G., Moore T.J.T., 
Ridge N.A., 2003, MNRAS, 340, 799 }
\mnref{McKee, C.F., Tan, J.C., 2003, \apj, 585, 850 }
\mnref{Martins F., Schaerer D., Hillier D.J., 2005, A\&A, 436, 1049 }
\mnref{Minniti D., et al., 2010, NewA, 15, 433 }

\mnref{Mizuno D.R., et al., 2008, PASP, 120, 1028 }
\mnref{Molinari S., et al., 2010, PASP, 122, 314 }
\mnref{Moore, T.J.T., 
Urquhart, J.S., Morgan, L.K., Thompson, M.A.\ 2012, \mnras, 426, 701 }
\mnref{Mottram J.C., Hoare M.G., Lumsden S.L., Oudmaijer R.D., Urquhart J.S., 
Sheret T.L., Clarke A.J., Allsopp J., 2007, A\&A, 476, 1019}
\mnref{Mottram J.C., Hoare M.G., Lumsden S.L., Oudmaijer R.D., Urquhart J.S.,
  Meade M.R., Moore T.J.T., Stead J.J., 2010, A\&A, 510, A89 }
\mnref{Mottram J.C., et al., 2011a, A\&A, 525, A149 }
\mnref{Mottram J.C., et al., 2011b, ApJ, 730, L33 }
\mnref{Ortiz, R., Lorenz-Martins, S., Maciel, W.J., \& Rangel, E.M.\ 2005,
  \aap, 431, 565 } 
\mnref{Pauldrach A.W.A., Hoffmann T.L., Lennon M., 2001, A\&A, 375, 161}
\mnref{Price S.D., Egan M.P., Carey S.J., Mizuno D.R., Kuchar T.A.,
	 2001, AJ, 121, 2819}
\mnref{Purcell C.R., et al., 2013, ApJS, 205, 1 }
\mnref{Reid M.J., et al., 2009, ApJ, 700, 137 }
\mnref{Rygl, K.L.J., Brunthaler, A.,
Reid, M.J., et al.\ 2010, \aap, 511, A2 } 
\mnref{Schuller, F., Menten, K.M., Contreras, Y., et al.\ 2009, \aap, 504, 415 
}
\mnref{Shu F.H., Adams F.C., Lizano S., 1987, ARA\&A, 25, 23 }
\mnref{Skrutskie M.F. et al., 2006, AJ, 131, 1163 }
\mnref{Sridharan T.K., Beuther H., Schilke P., 
Menten K.M., Wyrowski F., 2002, ApJ, 566, 931 }
\mnref{Thompson, M.A., Urquhart, J.S., Moore, T.J.T., Morgan, L.K.\ 
2012, \mnras, 421, 408 }
\mnref{Urquhart J.S., et al., 2007a, A\&A, 474, 891} 
\mnref{Urquhart J.S., Busfield A.L., Hoare M.G., Lumsden S.L., Clarke A.J., 
Moore T.J.T., Mottram J.C., Oudmaijer R.D., 2007b, A\&A, 461, 11 }
\mnref{Urquhart, J.S., Busfield, A.L., Hoare, M.G., et al.\ 2008, \aap, 487,
  253 } 
\mnref{Urquhart J.S., et al., 2009, A\&A, 501, 539 }

\mnref{Urquhart J.S., et al., 2011a, MNRAS, 410, 1237 }
\mnref{Urquhart, J.S., 
Morgan, L.K., Figura, C.C., et al.\ 2011, \mnras, 418, 1689 }
\mnref{Urquhart, J.S., Hoare, M.G., Lumsden, S.L., et al.\ 2012, 
\mnras, 420, 1656 }
\mnref{Urquhart, J.S., Moore, T.J.T., Schuller, F., et al.\ 2013a, MNRAS, 431,
  1752 }
\mnref{Urquhart, J.S., Thompson, M.A., Moore, T.J.T., et al.\ 2013b, MNRAS,
  in press (arXiv:1307.4105) }
\mnref{van Loo, S., Falle, 
S.A.E.G., \& Hartquist, T.W.\ 2007, \mnras, 376, 779 
}
\mnref{Wright E.L., et al., 2010, AJ, 140, 1868 }

\end{refs}

\clearpage
\begin{sidewaystable}
\footnotesize
\vspace*{-85mm}\hspace*{25mm}
\scalebox{0.88}{
\setlength{\tabcolsep}{1pt}

\begin{tabular}{lccrrl@{\hskip 5mm}lccrrl}
\hline
  \multicolumn{1}{c}{Name} &
  \multicolumn{1}{c}{RA} &
  \multicolumn{1}{c}{Dec} &
  \multicolumn{1}{c}{$D$} &
  \multicolumn{1}{c}{$L_{bol}$} &
  \multicolumn{1}{c}{Other} &
  \multicolumn{1}{c}{Name} &
  \multicolumn{1}{c}{RA} &
  \multicolumn{1}{c}{Dec} &
  \multicolumn{1}{c}{$D$} &
  \multicolumn{1}{c}{$L_{bol}$} &
  \multicolumn{1}{c}{Other} \\

  \multicolumn{1}{c}{} &
  \multicolumn{1}{c}{} &
  \multicolumn{1}{c}{} &
  \multicolumn{1}{c}{(kpc)} &
  \multicolumn{1}{c}{(\Lsolar)} &
  \multicolumn{1}{c}{Name} &
  \multicolumn{1}{c}{} &
  \multicolumn{1}{c}{} &
  \multicolumn{1}{c}{} &
  \multicolumn{1}{c}{(kpc)} &
  \multicolumn{1}{c}{(\Lsolar)} &
  \multicolumn{1}{c}{Name} \\
\hline
  G010.3208-00.1570B &18:09:01.48 &-20:05:08.0&  3.5      &41600   & &                        
  G010.8411-02.5919  &18:19:12.10 &-20:47:30.9&  1.9  &    23700  &GGD 27\\                  
  G012.0260-00.0317  &18:12:01.89 &-18:31:55.8& 11.1      &24600   &IRAS 18090-1832&          
  G012.1993-00.0342B &18:12:23.43 &-18:22:51.0& 12.0  &    34700  &IRAS 18094-1823\\         
  G012.9090-00.2607  &18:14:39.56 &-17:52:02.3&  2.4      &21700   &W33A&                     
  G017.6380+00.1566  &18:22:26.38 &-13:30:12.0&  2.2  &    53100  &GL2136\\                  
  G017.9789+00.2335A &18:22:49.14 &-13:10:01.5& 14.4      &31300   &&                         
  G018.3412+01.7681  &18:17:58.11 &-12:07:24.8&  2.8  &    21800  &IRAS 18151-1208\\         
  G020.7617-00.0638B &18:29:12.11 &-10:50:36.2& 11.8      &20800   &&                         
  G021.5624-00.0329  &18:30:36.07 &-10:07:11.1&  9.7  &    23700  &\\                        
  G023.3891+00.1851  &18:33:14.32 &-08:23:57.5&  4.5      &41900   &&                         
  G026.2020+00.2262  &18:38:18.51 &-05:52:57.5&  7.5  &    30500  &\\                        
  G027.1852-00.0812A &18:41:13.18 &-05:09:01.0& 13.0      &94300   &IRAS 18385-0512&          
  G028.3046-00.3871A &18:44:21.97 &-04:17:39.5& 10.0  &    38500  &\\                        
  G028.8621+00.0657  &18:43:46.25 &-03:35:29.3&  7.4     &146200   &IRAS 18411-0338&          
  G029.8620-00.0444  &18:45:59.55 &-02:45:06.5&  7.3  &    56000  &CH3OH 029.86-00.05\\      
  G030.1981-00.1691  &18:47:03.07 &-02:30:36.1&  7.3      &33200   &&                         
  G030.9585+00.0862B &18:47:31.83 &-01:42:59.6& 11.7  &    50400  &\\                        
  G030.9727+00.5620  &18:45:51.69 &-01:29:13.0& 12.6      &22900   &&                         
  G032.0451+00.0589  &18:49:36.56 &-00:45:45.5&  4.9  &    20400  &IRAS 18470-0049\\         
  G032.9957+00.0415A &18:51:24.45 &+00:04:34.1&  9.2      &22500   &IRAS 18488+0000&          
  G034.0126-00.2832  &18:54:25.06 &+00:49:56.6& 12.9  &    33800  &\\                        
  G034.0500-00.2977  &18:54:32.30 &+00:51:32.9& 12.9      &22500   &&                         
  G035.1979-00.7427  &18:58:13.00 &+01:40:31.2&  2.2  &    30900  &G35.2N\\                  
  G037.5536+00.2008  &18:59:09.95 &+04:12:15.7&  6.7      &38000   &IRAS 18566+0408&          
  G042.0341+00.1905A &19:07:28.20 &+08:10:53.3& 11.1  &    29300  &\\                        
  G042.0977+00.3521A &19:07:00.51 &+08:18:44.1& 10.9      &31300   &IRAS 19045+0813&          
  G042.0977+00.3521B &19:07:00.52 &+08:18:45.6& 10.9  &    31300  &IRAS 19045+0813\\         
  G042.1099-00.4466$^*$  &19:09:53.57 &+07:57:14.5&  8.7  &43400  &IRAS 19074+0752&      
  G043.0884-00.0109  &19:10:09.55 &+09:01:26.7& 11.1  &    32900  &\\                        
  G045.4543+00.0600B &19:14:21.27 &+11:09:15.5&  7.3      &34900   &IRAS 19120+1103&          
  G045.4543+00.0600C &19:14:21.24 &+11:09:20.2&  7.3  &    34900  &IRAS 19120+1103\\         
  G053.6185+00.0376  &19:30:23.04 &+18:20:26.6&  7.9      &20000   &&                         
  G060.5750-00.1861  &19:45:52.50 &+24:17:42.8&  7.5  &    30100  &IRAS 19437+2410\\         
  G060.8828-00.1295B$^*$ &19:46:20.14 &+24:35:29.3&  2.2  &21700 &S87 IRS1&             
  G062.5748+02.3875  &19:40:21.52 &+27:18:43.7& 13.4  &    96100  &IRAS 19383+2711\\         
  G064.8131+00.1743  &19:54:05.86 &+28:07:40.6&  8.2     &184300   &IRAS 19520+2759 &         
  G073.6525+00.1944  &20:16:21.96 &+35:36:06.3& 11.2 &    259200  &IRAS 20144+3526\\         
  G075.6014+01.6394  &20:15:48.16 &+38:01:31.3& 11.2      &28600  &&                         
  G076.3829-00.6210$^*$  &20:27:26.77 &+37:22:47.8&  1.4  &39700  &S106 IRS4 \\          
  G078.7641+01.6862  &20:24:51.67 &+40:39:25.3& 10.5      &42300   &&                         
  G078.8867+00.7087  &20:29:24.87 &+40:11:19.4&  3.3 &    185300  &GL2591\\                  
  G085.4102+00.0032A &20:54:14.36 &+44:54:04.6&  5.5      &20400   &&                         
  G090.2095+02.0405  &21:03:41.76 &+49:51:47.1&  7.4  &    29700  &IRAS 21020+4939 \\        
  G094.4637-00.8043  &21:35:09.11 &+50:53:09.6&  4.9      &20800   &IRAS 21334+5039&          
  G094.6028-01.7966  &21:39:58.25 &+50:14:20.9&  4.9  &    43200  &V645 Cyg  \\              
  G096.5438+01.3592  &21:35:43.82 &+53:53:09.4&  7.0      &22600   &IRAS 21340+5339 &         
  G097.5268+03.1837B &21:32:11.30 &+55:53:39.9&  6.9  &    30600  &\\                        
  G097.5268+03.1837C &21:32:10.69 &+55:53:35.4&  6.9      &21700   &S128 IRS 2&               
  G102.3533+03.6360  &21:57:25.19 &+59:21:56.7&  8.4 &    107000  &CPM 36\\                  
  G110.1082+00.0473B$^*$ &23:05:10.15 &+60:14:42.8&  4.3  &28400&S156A&                
  G111.2348-01.2385  &23:17:21.02 &+59:28:48.0&  4.4  &41900  &IRAS 23151+5912\\         
  G111.2824-00.6639B &23:16:03.85 &+60:01:56.8&  3.5  &24700       &IRAS 23138+5945&          
  G111.5671+00.7517  &23:14:01.76 &+61:27:19.9&  2.7  &44600     &NGC 7538 IRS 9\\          
  G133.6945+01.2166A &02:25:30.99 &+62:06:21.0&  2.0  &28500       &W3 IRS4 (part) &          
  G133.7150+01.2155  &02:25:40.78 &+62:05:52.5&  2.0  &206300     &W3 IRS5 \\                
  G135.2774+02.7981  &02:43:28.65 &+62:57:08.7&  6.0  &28500       &IRAS 02395+6244 &         
  G151.6120-00.4575  &04:10:11.86 &+50:59:54.5&  6.4  &21500     &CPM 12\\                  
  G192.6005-00.0479  &06:12:54.01 &+17:59:23.1&  2.0  &35600       &S255 IRS1&                
  G196.4542-01.6777  &06:14:37.06 &+13:49:36.5&  5.3  &94000     &IRAS 06117+1350 \\        
  G269.8539-00.0630  &09:11:08.34 &-48:15:56.3&  8.4  &27600       &&                         
  G274.0649-01.1460A &09:24:42.55 &-52:01:50.6&  5.7  &24300     &\\                        
  G281.0472-01.5432  &09:59:15.88 &-56:54:39.3&  7.0  &145100     &&                         
  G281.7578-02.0132  &10:01:21.58 &-57:42:56.4&  7.0  &32300     &\\                        
  G282.0598-00.5721  &10:09:26.56 &-56:43:49.5&  4.9  &20000       &&                         
  G282.8969-01.2727  &10:11:31.60 &-57:47:03.7&  7.0  &31900     &\\                        
  G289.9446-00.8909A &11:01:10.67 &-60:57:08.4&  8.3  &26200       &&                         
  G299.5265+00.1478  &12:21:50.64 &-62:31:42.4&  7.5  &34700     &\\                        
  G300.5047-00.1745A &12:30:03.60 &-62:56:48.4&  8.9  &42700      &&                         
  G301.8147+00.7808A &12:41:53.87 &-62:04:14.6&  4.4  &22000     &IRAS 12389-6147\\         
  G303.9973+00.2800$^*$  &13:00:41.62 &-62:34:20.8& 11.4  &20400 &&                     
  G304.3674-00.3359A &13:04:09.87 &-63:10:20.1& 11.8  &88200&   \\                        
  G304.6668-00.9654$^*$  &13:07:08.37 &-63:47:02.8& 11.4  &24800 &&                     
  G305.2017+00.2072A &13:11:10.45 &-62:34:38.6&  4.0  &30300&   CH3OH 305.20+00.21\\      
  G305.3676+00.2095  &13:12:36.49 &-62:33:32.3&  4.0  &28100       &&                         
  G305.5610+00.0124  &13:14:26.37 &-62:44:30.5&  4.0  &42000     &\\                        
  G308.9176+00.1231A &13:43:01.70 &-62:08:51.2&  5.3  &186800     &OH 308.918+00.123&        
  G309.9206+00.4790B &13:50:42.34 &-61:35:07.9&  5.4  &26600     &IRAS 13471-6120  \\       
  G310.0135+00.3892  &13:51:37.86 &-61:39:07.5&  3.2  &54600       &IRAS 13481-6124&          
  G319.3993-00.0135C &15:03:17.68 &-58:36:14.8& 11.7  &109800& \\                        
  G319.8366-00.1963  &15:06:54.49 &-58:32:58.8& 11.7  &38900       &&                         
  G321.0523-00.5070  &15:16:06.11 &-58:11:41.8&  9.1  &74400     &\\                        
  G321.3824-00.2861  &15:17:20.21 &-57:50:00.3&  9.4  &24600       &&                         
  G327.1192+00.5103  &15:47:32.81 &-53:52:39.4&  4.9  &41600     &CH3OH 327.120+00.511\\    
  G328.2523-00.5320A &15:57:59.83 &-53:58:00.5&  2.9  &40500       &CH3OH 328.25-00.53 &      
  G328.2523-00.5320B &15:57:59.38 &-53:57:57.4&  2.9  &21300     &\\                        
  G329.0663-00.3081  &16:01:09.93 &-53:16:02.3& 11.6  &65600       &CH3OH 329.07-00.31&       
  G331.2759-00.1891B &16:11:26.00 &-51:41:57.0&  4.9  &35000     &CH3OH 331.278-00.188\\    
  G331.3576+01.0626  &16:06:25.78 &-50:43:22.0&  4.5  &22200       &IRAS 16026-5035  &        
  G331.5131-00.1020  &16:12:09.96 &-51:28:37.1&  5.0  &69300     &OH 331.512-00.103\\       
  G331.5180-00.0947A &16:12:08.95 &-51:28:02.3&  5.0  &32100       &&                         
  G331.7953-00.0979  &16:13:28.04 &-51:16:46.8& 14.5  &105300   &\\                        
  G332.0939-00.4206  &16:16:16.47 &-51:18:25.2&  3.6  &76100      &&                         
  G332.8256-00.5498A$^*$ &16:20:11.07 &-50:53:16.2&  3.6 &207700 &\\                    
  G332.9868-00.4871  &16:20:37.81 &-50:43:49.7&  3.6  &26700       &&                         
  G333.1256-00.4367  &16:21:02.66 &-50:35:55.4&  3.6  &85000     &CH3OH 333.126-00.440\\    
  G334.8438+00.2095A &16:25:40.51 &-48:55:16.2& 10.6  &25400       &&                         
  G336.8308-00.3752  &16:36:26.16 &-47:52:30.9& 13.5  &50900     &\\                        
  G338.0008-00.1498A &16:40:04.02 &-46:51:18.1& 11.4  &50400       &&                         
  G338.2253-00.5094  &16:42:30.98 &-46:55:22.6& 13.7  &103000   &\\                        
  G338.3597+00.1430A &16:40:11.88 &-46:23:27.2& 12.8  &30100       &&                         
  G338.4712+00.2871  &16:39:58.91 &-46:12:36.5& 13.1  &86100     &IRAS 16363-4606\\         
  G338.4763+00.0418A &16:41:04.46 &-46:22:18.8& 12.6  &28200       &&                         
  G338.9196+00.5495  &16:40:34.05 &-45:42:08.0&  4.2  &32000     &\\                        
  G339.3316+00.0964  &16:44:04.39 &-45:41:27.2& 13.1  &39600       &&                         
  G339.6221-00.1209  &16:46:06.00 &-45:36:43.9&  2.8  &23800     &\\                        
  G339.8838-01.2588  &16:52:04.66 &-46:08:33.6&  2.7  &63900       &IRAS 16484-4603 &         
  G339.9489-00.5401  &16:49:07.95 &-45:37:58.8& 10.5  &20800     &EGO G339.95-0.54 \\       
  G342.9583-00.3180  &16:58:48.56 &-43:09:32.5& 12.7  &62200       &&                         
  G343.1261-00.0623  &16:58:17.21 &-42:52:07.1&  2.8  &66100     &IRAS 16547-4247  \\       
  G344.4257+00.0451C &17:02:08.62 &-41:47:10.2&  4.7  &23100       &&                         
  G344.6608+00.3401  &17:01:41.02 &-41:24:48.1& 12.7  &20300     &\\                        
  G345.4938+01.4677  &16:59:41.61 &-40:03:43.4&  2.4  &154400     &IRAS 16562-3959&          
  G345.5043+00.3480  &17:04:22.87 &-40:44:23.5&  2.0  &23700     &CH3OH 345.50+00.35\\      
  G349.7215+00.1203A &17:18:11.22 &-37:28:24.6& 11.3  &65900           \\
%
\hline
\end{tabular}}
\caption{Complete list of the massive protostars with $L>20000$\Lsolar.
  Objects marked with a superscript $*$ are also the central stars of HII regions.}
\clearpage
\end{sidewaystable}

\end{document}